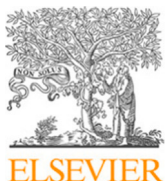

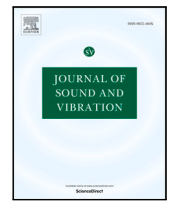

# A Kullback–Leibler divergence method for input–system–state identification

Marios Impraimakis

*University of Southampton, Southampton SO16 7QF, UK*

## A R T I C L E   I N F O

## A B S T R A C T

The capability of a novel Kullback–Leibler divergence method is examined herein within the Kalman filter framework to select the input–parameter–state estimation execution with the most plausible results. This identification suffers from the uncertainty related to obtaining different results from different initial parameter set guesses, and the examined approach uses the information gained from the data in going from the prior to the posterior distribution to address the issue. Firstly, the Kalman filter is performed for a number of different initial parameter sets providing the system input–parameter–state estimation. Secondly, the resulting posterior distributions are compared simultaneously to the initial prior distributions using the Kullback–Leibler divergence. Finally, the identification with the least Kullback–Leibler divergence is selected as the one with the most plausible results. Importantly, the method is shown to select the better performed identification in linear, nonlinear, and limited information applications, providing a powerful tool for system monitoring.

## 1. Introduction

Output-only system identification techniques have a long history of assessing the system condition when performed during their normal operation with ambient vibration data. In this direction, the stochastic modal identification techniques are introduced from output-only data combining high computational robustness efficiency with high estimation accuracy. To address the non-automated identification issue in output-only procedures extensive research is performed, and it is still ongoing. Rainieri and Fabbrocino [1] presented a literature review for the most common automated output-only dynamic identification techniques. Later, Pintelon et al. [2] introduced a novel multi-output algorithm for continuous-time operational modal analysis in the presence of harmonic disturbances with time-varying frequency. López-Aenlle et al. [3] proposed a procedure to optimize the mass-change strategy, which uses the modal parameters of the original system as the basic information. Reynders [4] examined the issue of data-driven state matrices derived using identified state sequences, while also investigated the uncertainty quantification. To provide insight on non-automated techniques, Priori et al. [5] studied a time domain output-only data-driven stochastic subspace identification technique for modal models providing also some user-defined parameters suggestions. Additionally, Grez et al. [6] developed a subspace identification method for the estimation of the structural parameters while rejecting the influence of the periodic input.

Online dynamic system monitoring, on the other hand, has been an active area of research for more than a half century for assessing the system condition in real-time. Over this period, the estimation of the dynamic states was firstly addressed mainly based on the Kalman filter and the particle filter [7–9]. While the state estimation has its applications, the original versions of these methods do not provide information on the remaining quantities of the system such as on the parameter and/or the input. This

---

deficiency is problematic for damage detection and the impracticability of needing to know the input in some circumstances. This has led to either the dynamic states and the parameters estimation in real-time, or the estimation of the dynamic states and the input in real-time. Methodologies are proposed for the joint parameter-state estimation [10,11] based on nonlinear filters such as the extended Kalman filter [12–15], the unscented Kalman filter [16–18], or the particle filter [19–21]. Nonetheless, the required input cannot always be measured, or the measurement of the input may be more unreliable than what is demanded. For instance, there is not a reliable means of accurately measuring the traffic and wind load on large structural systems. Methodologies for the joint input-state estimation or for unknown input dynamic state estimation are proposed in [22–28]. However, for known parameters the requirements of damage detection are not met. Though the procedures are useful and practical for real systems, they fail to address the complete identification problem when all quantities are unknown. The aforementioned, with respect to some assumed system properties, lead to the need of extraction of as much information as possible using only output data.

To this end, the simultaneous estimation of the dynamic states, the system parameters, and the input attracts the attention of the dynamic system monitoring community [29]. Over this period, Naets et al. [30] proposed an extended Kalman filter with augmented states for the unknown inputs and parameters. Dertimanis et al. [31] suggested an observer which combines the dual and the unscented Kalman filter. Furthermore, Castiglione et al. [32] introduced a time-varying auto-regressive model for the unknown input. Maes et al. [33] proposed a linearization of the system model around the states. Lei et al. [34] suggested a recursive nonlinear least squares strategy for the unknown input. Additionally, Song [35] proposed a minimum variance unbiased filtering with direct feedthrough. Rogers et al. [36] proposed a Gaussian process latent force model which allows a flexible Bayesian prior to be placed over the unknown forcing signal. Huang et al. [37] investigated the modulated colored noise for the unknown inputs. Importantly, Teymouri et al. [38] developed a Bayesian expectation–maximization methodology for the noise calibration. Finally, Capalbo et al. [39] proposed a parametric model order reduction for the augmented Kalman filtering, and the research is still ongoing.

However, developing a joint input–parameter–state methodology incorporates, apart from the identification convergence, the selection of the correct or the most plausible results. The main challenge is in the presence of uncertainty related to the different results from different initial parameter set guesses. Especially, if these are not reasonably close to the true parameter values, the algorithms converge to wrong values, and the user is unable to select the correct results among several executions. Specifically, the algorithms converge to suboptimal results which, although partially reproduce the measured data, they do not provide the correct parameter estimates.

A way to address this challenge is examined here using a generalized real-time procedure which selects automatically the most plausible results. The Kullback–Leibler divergence [40] is therefore employed, termed also as relative entropy.

Importantly, the Kullback–Leibler divergence has already been examined to select the better performed model class in real-time identification problems [16,41,42]. It has also been used extensively in machine learning [43] and in the sensor configuration optimization [44–47] for structural health monitoring applications. In another application, it has been proposed to appropriately select the distributions on the extended space that includes atomic coordinates and collective variables [48].

The Kullback–Leibler divergence, specifically, is a type of statistical distance, namely a measure of how different a probability distribution is from a second one. In this work, it is interpreted as the larger its value, the greater the information gained from the data between the posterior and the prior distribution, indicating a more complex identification which is less favorable.

In the examined method, the estimates of the Kalman filter, especially the unscented Kalman filter (UKF) [49] and the residual-based Kalman filter (RKF) [50] are evaluated simultaneously using the Kullback–Leibler divergence. Subsequently, the identification with the least Kullback–Leibler divergence is selected as the most plausible one.

The proposed method for selecting the most plausible results among several executions should not be confused with the Bayesian model evidence algorithms for online estimation [51–54]. Those approaches calculate the evidence of each candidate model given the available measured data, and they finally select the simpler ones over the unnecessarily complicated ones. The importance of those methods is highlighted by the fact that a more complicated model fits the data better than one which has fewer adjustable uncertain parameters, but it is likely results in data over-fitting and poor future predictions. This is attributed to the parameter fitting which depend too much on the detail of the data and the measurement noise. On the other hand, the proposed method solely solves the system identification problem for each candidate model.

The work is organized as follows: the Kullback–Leibler divergence for parameter set selection is developed in Section 2. In Section 3, the unknown input UKF is extended by the Kullback–Leibler divergence approach. In Section 4, the unknown input RKF is extended by the Kullback–Leibler divergence approach. Section 5 provides the summary and the detailed algorithmic tables. Importantly, Section 6, Section 7, and Section 8 investigate numerical applications on linear, limited information, and nonlinear systems. Investigations are also provided for various levels of observation noise and for comparison with systems with larger degrees of freedom (DOFs). Finally, further discussion, investigation of algorithmic parameters, and suggestions for future research is provided in Section 9. Last but not least, the conclusions are provided in Section 10.

## 2. The Kullback–Leibler divergence for parameter set selection

To select the initial parameter set $S_i$ in a Bayesian framework, one needs to use their prior probability distribution, and then assess their posterior probability plausibility. Let $\mathbf{S}$ be the space of the parameter sets $S_{i:i_{max}}$. The posterior probability $P(S_i \,|\, \mathbf{y}, \mathbf{S})$ of the parameter set $S_i$ is defined using the Bayes theorem as:

$$P(S_i \,|\, \mathbf{y}, \mathbf{S}) = \frac{p(\mathbf{y} \,|\, S_i) \cdot P(S_i \,|\, \mathbf{S})}{p(\mathbf{y} \,|\, \mathbf{S})} \tag{1}$$

where, $P(S_i \mid \mathbf{S})$ is the prior probability of $S_i$, $\mathbf{y}$ is the measurement vector, and $p(\mathbf{y} \mid S_i)$ is the evidence given the set $S_i$. The denominator is replaced by the summation of the prior probability and the likelihood for every parameter set, written as:

$$P(S_i \mid \mathbf{y}, \mathbf{S}) = \frac{p(\mathbf{y} \mid S_i) \cdot P(S_i \mid \mathbf{S})}{\sum_{i}^{i_{max}} \left\{ p(\mathbf{y} \mid S_i) \cdot P(S_i \mid \mathbf{S}) \right\}} \tag{2}$$

Let $\theta_j \in S_i$ be the parameter $j$ of the set $S_i$. The posterior probability distribution $p(\theta_j \mid \mathbf{y}, S_i)$ of $\theta_j$ is written as:

$$p(\theta_j \mid \mathbf{y}, S_i) = \frac{p(\mathbf{y} \mid \theta_j, S_i) \cdot p(\theta_j \mid S_i)}{\int_{\boldsymbol{\theta}} p(\mathbf{y} \mid \theta_j, S_i) \cdot p(\theta_j \mid S_i) \, \boldsymbol{d\theta}} = \frac{p(\mathbf{y} \mid \theta_j, S_i) \cdot p(\theta_j \mid S_i)}{p(\mathbf{y} \mid S_i)} \tag{3}$$

where, $p(\mathbf{y} \mid \theta_j, S_i)$ is the likelihood given the parameter $\theta_j$ and the set $S_i$, and $p(\theta_j \mid S_i)$ is the prior probability density function of $\theta_j$ given the set $S_i$. Here, computing the evidence $p(\mathbf{y} \mid S_i)$ for each set $S_i$ is not trivial. Specifically, the high-dimensional integral is usually analytically intractable, for instance when nonconjugate prior probabilities and/or latent variables exist.

To this end, stochastic simulation methods are used. Particularly, the Markov chain Monte Carlo methods generate samples from the posterior distribution, and then compute the likelihood using the following identity of a rearranged Bayes theorem for every $\theta_j$:

$$ln\big(p(\mathbf{y} \mid S_i)\big) = ln\big(p(\mathbf{y} \mid \theta_j, S_i)\big) + ln\big(p(\theta_j \mid S_i)\big) - ln\big(p(\theta_j \mid \mathbf{y}, S_i)\big) \tag{4}$$

where, the natural logarithm $ln(\bullet)$ is applied to avoid numerical overflows. Eq. (4) is also written as [55]:

$$ln\big(p(\mathbf{y} \mid S_i)\big) = \int_{\boldsymbol{\theta}} ln\big(p(\mathbf{y} \mid \theta_j, S_i)\big) \, p(\theta_j \mid \mathbf{y}, S_i) \, \boldsymbol{d\theta} - \int_{\boldsymbol{\theta}} ln\left( \frac{p(\theta_j \mid \mathbf{y}, S_i)}{p(\theta_j \mid S_i)} \right) p(\theta_j \mid \mathbf{y}, S_i) \, \boldsymbol{d\theta} \tag{5}$$

where, the first expectation term measures the posterior average data fit of the parameter set $S_i$, while the penalty-type second one represents the Kullback–Leibler divergence $D_{KL}$ between the parameter posterior and prior probability distributions.

Specifically, the Kullback–Leibler divergence measures the similarity between $q_I(\boldsymbol{\theta}) = p(\theta_j \mid S_i)$ and $q_{II}(\boldsymbol{\theta}) = p(\theta_j \mid \mathbf{y}, S_i)$ as:

$$D_{KL}\big(\mathbf{q_{II}}(\boldsymbol{\theta}) \,\big\|\, \mathbf{q_I}(\boldsymbol{\theta})\big) := \int_{-\infty}^{\infty} \mathbf{q_{II}}(\boldsymbol{\theta}) \cdot ln\left( \frac{\mathbf{q_{II}}(\boldsymbol{\theta})}{\mathbf{q_I}(\boldsymbol{\theta})} \right) \boldsymbol{d\theta} \tag{6}$$

where the following property applies,

$$D_{KL}\big(\mathbf{q_{II}}(\boldsymbol{\theta}) \,\big\|\, \mathbf{q_I}(\boldsymbol{\theta})\big) \begin{cases} \geq 0 & \forall \quad q_I(\boldsymbol{\theta}), \, q_{II}(\boldsymbol{\theta}) \\ = 0 & \text{when } q_I(\boldsymbol{\theta}) = q_{II}(\boldsymbol{\theta}) \end{cases} \tag{7}$$

Finally, for Gaussian distributions of $d$ dimension, it is finally written as:

$$D_{KL}\big(\mathbf{q_{II}}(\boldsymbol{\theta}) \,\big\|\, \mathbf{q_I}(\boldsymbol{\theta})\big) = \frac{1}{2} \left[ ln\left( \frac{\mathbf{W_I}}{\mathbf{W_{II}}} \right) - d + trace\left( \mathbf{W_I^{-1}} \cdot \mathbf{W_{II}} \right) + \left( \mathbf{E_I} - \mathbf{E_{II}} \right)^T \cdot \mathbf{W_I^{-1}} \cdot \left( \mathbf{E_I} - \mathbf{E_{II}} \right) \right] \tag{8}$$

where, $\mathbf{E}$ and $\mathbf{W}$ are the mean and variance values of the prior I and the posterior II distribution, respectively. This Kullback–Leibler divergence representation allows for online and real-time evaluation of the identification with each parameter set $S_i$, and importantly, allows for a direct comparison of a number of simultaneous identifications from several sets.

## 3. Input–parameter–state estimation using the Kullback–Leibler divergence the unscented Kalman filter

For the mathematical implementation of the unknown input UKF [49] consider the nonlinear process equation in the continuous-time and the state-space format:

$$\dot{\mathbf{z}}(t) = f\big(\mathbf{z}(t), \mathbf{u}(t)\big) + \boldsymbol{v}(t) \tag{9}$$

and the nonlinear observation equation:

$$\mathbf{y}(t) = h\big(\mathbf{z}(t), \mathbf{u}(t)\big) + \boldsymbol{\eta}(t) \tag{10}$$

where, $\mathbf{y}(t)$ is the observation vector. The state vector $\mathbf{z}(t) = [\mathbf{x}(t), \dot{\mathbf{x}}(t), \boldsymbol{\theta}]^T$ includes the dynamic states and the system parameters. Also, $f(\bullet)$ and $h(\bullet)$ are the state transition function and the observation function, respectively, which take into account the, unknown here, input vector $\mathbf{u}(t)$. Lastly, $\boldsymbol{v}(t)$ and $\boldsymbol{\eta}(t)$ are the process and the measurement noise of covariance matrices $\mathbf{Q}(t)$ and $\mathbf{R}(t)$, respectively. Eqs. (9) and (10) are discretized as:

$$\mathbf{z_k} = F(\mathbf{z_{k-1}}, \mathbf{u_{k-1}}) + \boldsymbol{v_{k-1}} \tag{11}$$

and,

$$\mathbf{y_k} = h(\mathbf{z_k}, \mathbf{u_k}) + \boldsymbol{\eta_k} \tag{12}$$

where,

$$F(\mathbf{z_{k-1}}, \mathbf{u_{k-1}}) = \mathbf{z_{k-1}} + \int_{(k-1)\Delta t}^{(k)\Delta t} f\big(\mathbf{z}(t), \mathbf{u}(t)\big) \, dt \tag{13}$$

Note that $k$ refers to $k\Delta t$ time instant, where $\Delta t$ is the sampling period. The discretized process and observation covariance matrices are:

$$\mathbf{Q_{k-1}} \approx \frac{\mathbf{Q}\big((k-1)\Delta t\big)}{\Delta t}, \quad \mathbf{R_k} \approx \frac{\mathbf{R}(k\Delta t)}{\Delta t} \tag{14}$$

It is assumed, though, that the matrices are constant during the whole process, where being constant does not harm the estimation success; an investigation of their exact value, which importantly highly affects the success of the estimation [56], is shown in Section 9.

Simultaneously estimating the unknown input is possible using the UKF. Employing the predicted states at time step $k$, the input is estimated using the continuous equation of motion at the time instant $k\Delta t$ as:

$$\mathbf{u_k^p} \approx \mathbf{G}\big(\bar{\mathbf{x}}_k^m, \dot{\mathbf{x}}_k^p, \mathbf{x}_k^p, \boldsymbol{\theta}_k^p\big) \tag{15}$$

where, $\mathbf{G}(\bullet)$ is the linear or nonlinear system model, which contains the estimated parameters and, thus, it is updated in every step. Importantly, the predicted states are estimated using Eq. (11); with the prior input only.

Subsequently, the potential zero or non-zero valued known inputs replace the related $\mathbf{u_k^p}$ rows. The resulting predicted input is used in the updating process with the measurements in Eq. (12).

However, this predicted $\mathbf{u_k^p}$ input is erroneous since neither the predicted states have been updated with the measurements yet, nor the system parameters [49]. This is generally acceptable and its noise characteristics are incorporated into the measurement noise $\boldsymbol{\eta}_k$ of the observation equation. Here, the input error is usually larger than the state errors, and ignoring to update it would yield large estimation errors steamed from the misrepresentation of the measurement noise covariance matrix that leads to unrealistic gains for the Kalman-types filters. The next steps address this issue.

The input estimation is corrected using the measurements dynamic states and parameters as:

$$\mathbf{u_k^e} \approx \mathbf{G}\big(\bar{\mathbf{x}}_k^m, \dot{\mathbf{x}}_k, \mathbf{x}_k, \boldsymbol{\theta}_k\big) \tag{16}$$

and then, the known input rows replace the related rows of the final $\mathbf{u_k^e}$.

The measurement error of the accelerations still exists in contrast to the displacement, the velocity, and the parameter error. However, this process error is filtered when modeled by $\boldsymbol{\eta}_k$. The final $\mathbf{u_k^e}$ is successively used at the prediction calculation of the next step $k+1$, and the overall procedure is repeated for a number of examined initial parameter sets.

The developed process is successively implemented for all steps, and the overall procedure is repeated for a number of examined initial parameter sets. To this end, $D_{KL}$ evaluates the current parameter set using Eq. (8), where here, $\mathbf{E_I}$, $\mathbf{E_{II}}$, $\mathbf{W_I}$, and $\mathbf{W_{II}}$ are the mean values and the covariance matrices, respectively, of the initial parameter set denoted by I (not the current step prior), and the posterior parameter estimates denoted by II. Importantly, the posterior parameter estimates are provided online within the unscented Kalman filter process. Subsequently, the evaluation is performed for all parameter sets in a simultaneous fashion. Finally, the identification with the least $D_{KL}$ provides the most plausible input-parameter-state estimation.

Assuming additive Gaussian noise with zero mean, without loss of generality, the procedure steps are provided in Table 1A of Section 5. In this pseudo-code, $\lambda$ is given by $\lambda = \alpha^2(L + \kappa) - L$ with secondary parameter $\kappa = 0$ or $3 - L$ [17], where $L$ is the dimension of $\mathbf{z}$. The constant $\alpha \in [10^{-4}, 1]$ determines the spread of the sigma points around $\mathbf{z}$, while the weights $V$ are given by:

$$V_0^m = \frac{\lambda}{L + \lambda}$$

$$V_0^c = \frac{\lambda}{L + \lambda} + (1 - \alpha^2 + \beta) \tag{17}$$

$$V_i^m = V_i^c = \frac{\lambda}{2(L + \lambda)} \quad i = 1, \dots, 2L$$

where, $\beta$ is a constant that incorporates prior information of the $\mathbf{z}$ distribution.

## 4. Input–parameter–state estimation using the Kullback–Leibler divergence and the residual Kalman filter

For the mathematical implementation of the unknown input residual-based Kalman filter [50] consider the process equation in the continuous-time and the state-space format:

$$\dot{\mathbf{z}} = \mathbf{A}\,\mathbf{z} + \mathbf{B}\,\mathbf{u} \tag{18}$$

where, $\mathbf{A}(\boldsymbol{\theta})$ is the system matrix depended on the unknown parameter vector $\boldsymbol{\theta}$, and $\mathbf{B}$ is the input distribution matrix.

The discrete-time transformation of the system and the input matrices is provided by the zero-order hold assumption for the input in between the time instants $k\Delta t$, as:

$$\mathbf{A_d} = e^{\mathbf{A}\Delta t} \approx \mathbf{I_{2n \times 2n}} + \Delta t\,\mathbf{A} + \frac{\Delta t}{2}\mathbf{A}^2 \tag{19}$$

and,

$$\mathbf{B_d} = \int_0^{\Delta t} e^{\mathbf{A}\tau}\,\mathbf{B}\,d\tau = \mathbf{A}^{-1}\big[\mathbf{A_d} - \mathbf{I_{2n \times 2n}}\big]\mathbf{B} \approx \Delta t\mathbf{B} \tag{20}$$

The state-space model of Eq. (18) in the discrete-time, including the noise term $\mathbf{w}_k$, is written as:

$$\mathbf{z}_{k+1} = \mathbf{A_d}\,\mathbf{z}_k + \mathbf{B_d}\,\mathbf{u}_k^e + \mathbf{w}_k \tag{21}$$

where, $\mathbf{u}_k^e$ and $\mathbf{A_d}(\boldsymbol{\theta_k})$ are the estimated input and system matrix of the prior step which are considered as known quantities at the $k+1$ step.

The equation which relates the measurements $\mathbf{y}$ to the estimated dynamic states is written as:

$$\mathbf{y}_{k+1} = \mathbf{H}\,\mathbf{z}_{k+1} + \mathbf{w}_{k+1}^y \tag{22}$$

where, $\mathbf{H}$ is the observation matrix mapping the measurements to the dynamic states, and here is chosen to not depend on the unknown parameters and input. To this end and for limited information applications, $\mathbf{y}$ consists of displacement and velocity pseudo-measurements; the integrated of the actual acceleration measurements. Additionally, the accelerations which are not measured are assumed to be equal to the estimated accelerations of the previous step. Specifically here, the matrix $\mathbf{H}$ is introduced as the observation matrix mapping measurements to dynamical states. This matrix only accommodates displacements and velocities observed from all DOFs with real or pseudo-measurements.

It may seem here that the acceleration responses are not covered by the observation matrix. However, this is chosen intentionally since it addresses two problems. Firstly, the unknown input and parameters have not yet been estimated for the step $k+1$, and secondly the prior step parameters and input possibly affect negatively the observation equation when they are inaccurate.

More importantly, the presented observation model reflects the model for the pseudo-measurements rather than the actual measured quantities. In that case, the actual observation model relating the observed quantities to the state vector is not defined. To clarify how different measurement scenarios are accommodated within this approach and at which step they weigh in, the reader is referred to [50]. Additionally, a discussion is provided in Section 9 for the assumption of the unmeasured acceleration responses to be equal to the predicted accelerations from previous steps to legitimate and justify such an approach.

The predicted covariance matrix $\mathbf{P}_{k+1}$ of the dynamic states is then written as:

$$\mathbf{P}_{k+1} = \mathbf{A_d}\,\mathbf{P}_k\,\mathbf{A_d^T} + \mathbf{Q_{d(k)}} \tag{23}$$

where, the covariance matrices of the measurements and the system process are derived in similar manner to Section 3 using Eq. (14).

Having provided the posterior prediction model for the dynamic states and their covariances, the update process starts according to the Kalman filter. The updated dynamic state estimate is specifically derived by a correction of the predicted dynamic states using the measurement pre-fit residual, multiplied and controlled by the optimal Kalman gain $\mathbf{J}$ given as:

$$\mathbf{J}_{k+1} = \mathbf{P}_{k+1}\,\mathbf{H^T}\,\mathbf{N}_{k+1}^{-1} \tag{24}$$

where, the pre-fit residual covariance $\mathbf{N}$ is:

$$\mathbf{N}_{k+1} = \mathbf{H}\,\mathbf{P}_{k+1}\,\mathbf{H^T} + \mathbf{R_d} \tag{25}$$

The final estimation of the posterior dynamic states is then given by:

$$\mathbf{z}_{k+1} = \mathbf{z}_{k+1} + \mathbf{J}_{k+1}\,(\mathbf{y}_{k+1} - \mathbf{H}\,\mathbf{z}_{k+1}) \tag{26}$$

while the final estimation of the covariance of the dynamic states is given by:

$$\mathbf{P}_{k+1} = (\mathbf{I_{n \times n}} - \mathbf{J}_{k+1}\,\mathbf{H})\,\mathbf{P}_{k+1} \tag{27}$$

For Eqs. (26) and (27), the same quantity on the right and left hand side implies that they are re-calculated at the same time step; the a priori estimate of the right hand side is used for the calculation of a posteriori estimate on the left hand side.

Once the dynamic states are filtered using the pseudo-measurements, and with the use of the parameters of the prior step: The input at the current step is approximated by the system model at the time instant $(k+1)\,dt$ as:

$$\mathbf{u}_{k+1}^e \approx \mathbf{G}(\ddot{\mathbf{x}}_k^m,\,\mathbf{z}_k,\,\boldsymbol{\theta_k}) \tag{28}$$

where, $\mathbf{G}(\bullet)$ is the linear or nonlinear system model, which contains the prior step estimated parameters. Importantly, the predicted states are estimated using Eq. (26); with the prior input and parameters only. The known input rows of $\mathbf{u}_{k+1}^e$ are replaced by the potential known zero or non-zero valued inputs.

For the parameter estimation, a sensitivity analysis approach is implemented by the Taylor series expansion truncated after the linear term. To provide a real-time estimation specifically, the measured outputs are chosen to be accelerations instead of the modal parameters, written as:

$$\boldsymbol{\epsilon}_{k+1} = {}^{ma}_{k+1} - \mathbf{a}_{k+1} \approx \mathbf{r}_{k+1} + \mathbf{U}_{k+1}(\boldsymbol{\theta} - \boldsymbol{\theta}_{k+1}) \tag{29}$$

where, $\boldsymbol{\epsilon}_{k+1}$, ${}^{ma}_{k+1}$, and $\mathbf{a}_{k+1}$ denote the error, the acceleration measurements, and the predicted output, respectively, at the step $k+1$. The sensitivity matrix $\mathbf{U}_{k+1}$, which does not need an initial value or prior information, is written as:

$$\mathbf{U}_{k+1} = -\left[\frac{\partial\,{}^{ma}_{k+1}}{\partial\boldsymbol{\theta}}\right]_{\boldsymbol{\theta}=\boldsymbol{\theta}_{k+1}} \tag{30}$$

where, the error $\boldsymbol{\epsilon}_{k+1}$ is assumed to be small for the parameter vector $\boldsymbol{\theta}$ in the vicinity of $\boldsymbol{\theta}_{k+1}$.

At each step, Eq. (29) is solved by a Gauss–Newton gradient approach: The prior parameter estimates are corrected as:

$$\boldsymbol{\theta}_{k+1} = \boldsymbol{\theta}_k + \boldsymbol{\Delta\theta}_{k+1} \cdot e^{-\mu \, \|\boldsymbol{\rho}_{k+1}\|_2} \tag{31}$$

where, $\mu$ is a scaling parameter and $\|\boldsymbol{\rho}_{k+1}\|_2$ is the Euclidean norm of the residual of the system model estimation. In practice, $e^{-\mu \|\boldsymbol{\rho}_{k+1}\|_2}$ acts as a control factor for the convergence speed and fluctuation range. An investigation of this scaling parameter is shown in Section 9. A similar investigation can be done to define it for different types of model parameters within various dynamic systems.

For Eq. (31), the residual of the system model estimation is:

$$\boldsymbol{\rho}_{k+1} = \mathbf{u}_{k+1}^{\mathbf{e}} - \mathbf{G}\big(\ddot{\mathbf{x}}_{k+1}^{\mathbf{m}}, \mathbf{x}_{k+1}', \mathbf{x}_{k+1}, \boldsymbol{\theta}_k\big) \tag{32}$$

where, $\mathbf{u}_{k+1}^{\mathbf{e}}$ is the estimated input for the step $k+1$, and the dynamic states are provided by the Kalman filter.

For the objective function, the least square approach is formulated with an additional scaling parameter $\lambda^2$ balancing specifically the contribution of the parameter estimates. The final optimal $\boldsymbol{\Delta\theta}_{k+1}$ correction is provided by:

$$\boldsymbol{\Delta\theta}_{k+1} = [\mathbf{U}_{k+1}^{\mathrm{T}} \, \mathbf{U}_{k+1} + \lambda^2 \, \mathbf{I}]^{-1} \, \mathbf{U}_{k+1}^{\mathrm{T}} \, \boldsymbol{\rho}_{k+1} \tag{33}$$

where, $\lambda^2$ remains constant during the real-time procedure. An investigation of this scaling parameter is shown in Section 9. A similar investigation can be done to set both scaling parameters for different types of model parameters within various dynamic systems.

**Table 1**
Input–system–state identification using the Kullback–Leibler divergence.

| A. Kullback–Leibler divergence with UKF. | B. Kullback–Leibler divergence with RKF. |
|---|---|
| **step 1:** | **step 1:** |
| • $S = 1$   (Identification Set) | • $S = 1$   (Identification Set) |
| • $k = 0$   (Time step) | • $k = 0$   (Time step) |
| • $\mathbf{z}_k = \mathbb{E}[\mathbf{z}_0]$   ($\mathbb{E}$ stands for Expectation) | • $\mathbf{z}_k = \mathbb{E}[\mathbf{z}_0]$ |
| • $\mathbf{P}_k = \mathbb{E}[(\mathbf{z}_0 - \mathbf{z}_k)(\mathbf{z}_0 - \mathbf{z}_k)^T]$   (Covariance matrix) | • $\mathbf{P}(k) = \mathbb{E}[(\mathbf{z} - \mathbf{z}(k))(\mathbf{z} - \mathbf{z}(k))^T]$ |
| • $\mathbf{u}_k^{\mathbf{e}} = \mathbb{E}[\mathbf{u}_0^{\mathbf{e}}]$ | • $\theta(k) = \mathbb{E}[\theta]$ |
| | • $\lambda^2 \approx 5 \cdot 10^{-2}$ |
| | • $\mu \approx 5 \cdot 10^{-3}$ |
| | • $\mathbf{u}_k^{\mathbf{e}} = \mathbb{E}[\mathbf{u}_0^{\mathbf{e}}]$ |
| **step 2:** | **step 2:** |
| • $\mathbf{Z}_k = [\mathbf{z}_k - \sqrt{(L+\lambda)\mathbf{P}_k}, \ \ \mathbf{z}_k, \ \ \mathbf{z}_k + \sqrt{(L+\lambda)\mathbf{P}_k}]$ | |
| • $k = k + 1$ | |
| • $\mathbf{Z}_p = F(\mathbf{Z}_{k-1}, \mathbf{u}_{k-1}^{\mathbf{e}})$ (p stands for prediction) | • $\mathbf{z}(k+1) = \mathbf{A}_d \, \mathbf{z}(k) + \mathbf{B}_d \, \mathbf{u}^{\mathbf{e}}(k)$ |
| • $\mathbf{z}_p = \sum_{i=0}^{2L} V_i^m \mathbf{Z}_{i,p}$ | |
| • $\mathbf{u}_k^{\mathbf{p}} \approx \mathbf{G}(\ddot{\mathbf{x}}_k^{\mathbf{m}}, \mathbf{z}_p, \theta_k^{\mathbf{p}})$ ($G(\bullet)$ involves estimated parameters and it is updated in every step) | |
| • Replace all $\mathbf{u}_k^{\mathbf{p}}$ zero/non-zero known input rows | |
| • $\mathbf{P}_p = \sum_{i=0}^{2L} V_i^c \, [\mathbf{Z}_{i,p} - \mathbf{z}_p][\mathbf{Z}_{i,p} - \mathbf{z}_p]^T + \mathbf{Q}_{k-1}$ | • $\mathbf{P}(k+1) = \mathbf{A}_d \, \mathbf{P}(k) \, \mathbf{A}_d^{\mathsf{T}} + \mathbf{Q}_d$ |
| • $\mathbf{Y}_i = h(\mathbf{Z}_{i,p}, \mathbf{u}_k^{\mathbf{p}})$ | • $\mathbf{J}(k+1) = \mathbf{P}(k+1) \, \mathbf{H}^T \Big(\mathbf{R}_d + \mathbf{H} \, \mathbf{P}(k+1) \, \mathbf{H}^T\Big)^{-1}$ |
| • $\mathbf{y} = \sum_{i=0}^{2L} V_i^m \mathbf{Y}_i$ | |
| • $\mathbf{P}_m = \sum_{i=0}^{2L} V_i^c \, [\mathbf{Y}_i - \mathbf{y}][\mathbf{Y}_i - \mathbf{y}]^T + \mathbf{R}_k$ | |
| • $\mathbf{P}_s = \sum_{i=0}^{2L} V_i^c \, [\mathbf{Z}_{i,p} - \mathbf{z}_p][\mathbf{Y}_i - \mathbf{y}]^T$ | |
| If $\mathbf{y}_k^{\mathbf{m}}$ is the measurement vector at the current time step: | If $\mathbf{y}_k^{\mathbf{m}}$ is the measurement vector at the current time step: |
| | • Create the Pseudo-measurements (double integration [50]) |
| • $\mathbf{z}_k = \mathbf{z}_p + \mathbf{P}_s \, \mathbf{P}_m^{-1} \cdot (\mathbf{y}_k^{\mathbf{m}} - \mathbf{y})$   (State estimation) | • $\mathbf{z}(k+1) = \mathbf{z}(k+1) + \mathbf{J}(k+1) \cdot \big(^{\mathbf{m}}\mathbf{z}(k+1) - \mathbf{H} \, \mathbf{z}(k+1)\big)$ |
| • $\mathbf{P}_k = \mathbf{P}_p - \mathbf{P}_s \cdot (\mathbf{P}_s \, \mathbf{P}_m^{-1})^T$   (Covariance estimation) | • $\mathbf{P}(k+1) = \Big(\mathbf{I} - \mathbf{J}(k+1) \, \mathbf{H}\Big) \mathbf{P}(k+1)$ |
| • $\mathbf{u}_k^{\mathbf{e}} \approx \mathbf{G}(\ddot{\mathbf{x}}_k^{\mathbf{m}}, \mathbf{z}_k, \boldsymbol{\theta}_k)$   (Final Input estimation) | • $\mathbf{u}_{k+1}^{\mathbf{e}} \approx \mathbf{G}(\ddot{\mathbf{x}}_{k+1}^{\mathbf{m}}, \mathbf{z}_{k+1}, \boldsymbol{\theta}_k)$   (Final Input estimation) |
| • Replace all $\mathbf{u}_k^{\mathbf{e}}$ zero/non-zero known input rows | • Replace all $\mathbf{u}_{k+1}^{\mathbf{e}}$ zero/non-zero known input rows |
| | • $\mathbf{U}_{k+1} = -[\partial^{\mathbf{m}} \mathbf{a}_{k+1} / \partial \theta]_{\theta=\theta_{k+1}}$ |
| | • $\rho(k+1) = \mathbf{u}(k-1) - \mathbf{G}(\ddot{\mathbf{x}}_{k+1}^{\mathbf{m}}, \dot{\mathbf{x}}_{k+1}, \mathbf{x}_{k+1}, \boldsymbol{\theta}_k)$ |
| | • $\boldsymbol{\Delta\theta}(k+1) = [\mathbf{U}^T \mathbf{U} + \lambda^2 \mathbf{I}]^{-1} \, \mathbf{U}^T \, \rho(k+1)$ |
| | • $\theta(k+1) \ \ = \theta(k) + \boldsymbol{\Delta\theta}(k+1) \cdot e^{-\mu \|\rho(k+1)\|_2}$ |
| **step 3:** | **step 3:** |
| • $D_{KL}\big(\mathbf{q_{II}}(\theta) \big\| \mathbf{q_I}(\theta)\big) = \int_{-\infty}^{\infty} \mathbf{q_{II}}(\theta) \cdot ln(\mathbf{q_{II}}(\theta)/\mathbf{q_I}(\theta)) \, d\theta$ | • $D_{KL}\big(\mathbf{q_{II}}(\theta) \big\| \mathbf{q_I}(\theta)\big) = \int_{-\infty}^{\infty} \mathbf{q_{II}}(\theta) \cdot ln(\mathbf{q_{II}}(\theta)/\mathbf{q_I}(\theta)) \, d\theta$ |
| • Go to Step 2 until $k = k_{max}$ | • Go to Step 2 until $k = k_{max}$ |
| **step 4:** | **step 4:** |
| • Go to Step 1 until $S = S_{max}$ | • Go to Step 1 until $S = S_{max}$ |

Importantly, it is seen here that the transitional model assumed for the system parameters is involved in the full input–parameter–state estimation success, and taking partial derivatives is required. Also, the scaling factor is tied to the difference between the estimated and the predicted input forces. The nature, the order of magnitude, and the governing equations for the input and model parameters are different, but this approach shows to be beneficial in yielding stable estimates for the model parameters.

The developed process is successively implemented for all steps, and the overall procedure is repeated for a number of examined initial parameter sets. To this end, $D_{KL}$ evaluates the current parameter set using Eq. (8), where here, $\mathbf{E_I}$, $\mathbf{E_{II}}$, $\mathbf{W_I}$, and $\mathbf{W_{II}}$ are the mean values and the covariance matrices, respectively, of the initial parameter set denoted by I (not the current step prior), and the posterior estimates related to each parameter dynamic state denoted by II. Importantly, the posterior estimates are provided online within the residual-based Kalman filter process. Subsequently, the evaluation is performed for all parameter sets in a simultaneous fashion. Finally, the identification with the least $D_{KL}$ provides the most plausible input–parameter–state estimation. The detailed procedure steps of this procedure are provided in Table 1B of Section 5.

## 5. Procedure summary

The procedure is illustrated here where each step is detailed in Table 1:

1. **Initialize the identification**. Set the initial estimates for the dynamic states, the parameters, the input, and the algorithmic parameters.
2. **Execute the input–parameter–state estimation**. Predict the dynamic states using the discrete state-space model and the prior input and parameters. Create the pseudo-measurements by integrating the output data to provide the displacement and velocity pseudo-measurements, while use additional displacement sensing for highly nonlinear systems. The non-measured accelerations are replaced by the prior step acceleration estimates. Finally, estimate the current step dynamic states, the current step input using the updated system model, and the current step parameters. The algorithmic parameters mentioned here derive a reasonable value after Section 9 investigation. In the same section, the effect of taking the unmeasured acceleration responses equal to their estimated priors is discussed to provide insight into the convergence and stability of the estimates.
3. **Estimate the Kullback–Leibler divergence**. Estimate the Kullback–Leibler divergence for the examined initial parameter set using the Kalman filter mean and variance values compared to the initial ones. Repeat Step 2 and 3 for all time steps. The mean and variances values are specifically provided online within the Kalman filter procedure by $\mathbf{z}$ and $\mathbf{P}$.
4. **Select the final estimation**. Repeat Steps 1–3 for the next initial parameter set $S_{i+1}$. Select the most plausible input–parameter–state estimation using the identification with the least $D_{KL}$.

Importantly, the initial distribution assumption for the parameters to run the Kullback–Leibler divergence $\mathbf{W_I}$ is equal to the examined mean value of the initial parameter states with the unit covariance matrix; namely a Gaussian distribution is defined assuming that the user does not have any specific prior information. All applications in this work are using this assumption.

## 6. Application to linear systems

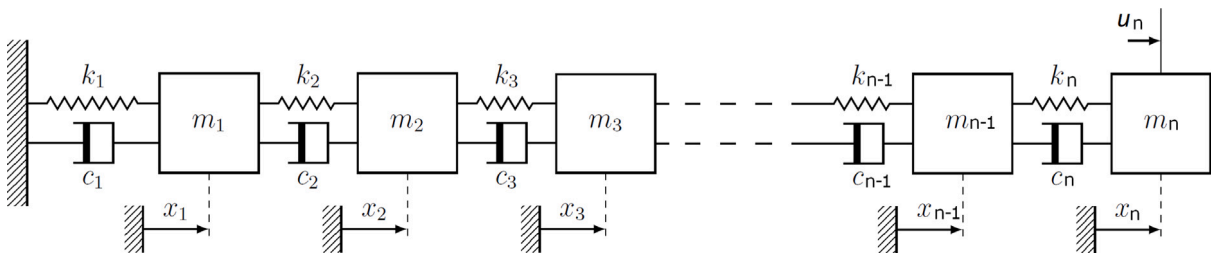

**Fig. 1.** Linear or nonlinear n-DOF system excited at DOF n, or at more DOFs.

For the linear numerical application consider a 3-DOF system based on Fig. 1, with the implementation of the unscented Kalman filter with the Kullback–Leibler divergence provided by Table 1A. The system is described by the following equation:

$$\mathbf{M}\begin{Bmatrix}\ddot{x}_1(t)\\ \ddot{x}_2(t)\\ \ddot{x}_3(t)\end{Bmatrix} + \mathbf{C}\begin{Bmatrix}\dot{x}_1(t)\\ \dot{x}_2(t)\\ \dot{x}_3(t)\end{Bmatrix} + \mathbf{K}\begin{Bmatrix}x_1(t)\\ x_2(t)\\ x_3(t)\end{Bmatrix} = \begin{Bmatrix}0\\ 0\\ u_3(t)\end{Bmatrix} \tag{34}$$

where the system matrices which need to be identified (apart from $\mathbf{M}$) are:

$$\mathbf{M} = \begin{bmatrix} m_1 & 0 & 0 \\ 0 & m_2 & 0 \\ 0 & 0 & m_3 \end{bmatrix} = \begin{bmatrix} 1 & 0 & 0 \\ 0 & 1 & 0 \\ 0 & 0 & 1 \end{bmatrix}, \quad \mathbf{C} = \begin{bmatrix} c_1 + c_2 & -c_2 & 0 \\ -c_2 & c_2 + c_3 & -c_3 \\ 0 & -c_3 & c_3 \end{bmatrix} = \begin{bmatrix} 0.25 + 0.5 & -0.5 & 0 \\ -0.5 & 0.5 + 0.75 & -0.75 \\ 0 & -0.75 & 0.75 \end{bmatrix},$$

$$\mathbf{K} = \begin{bmatrix} k_1 + k_2 & -k_2 & 0 \\ -k_2 & k_2 + k_3 & -k_3 \\ 0 & -k_3 & k_3 \end{bmatrix} = \begin{bmatrix} 9 + 11 & -11 & 0 \\ -11 & 11 + 13 & -13 \\ 0 & -13 & 13 \end{bmatrix}$$

$$(35)$$

with initial conditions $\mathbf{x}(0) = [0 \quad 0 \quad 0]^T$ and $\dot{\mathbf{x}}(0) = [0 \quad 0 \quad 0]^T$.

In order to create synthetic measurements, the Runge–Kutta 4th order method of integration is used to compute the system response for 30 s. The sampling frequency for the dynamic state measurements is considered to be 100 Hz, therefore the time discretization $\Delta t$ is 0.01 s. Finally, to consider measurement noise, each response signal is contaminated by a Gaussian white noise sequence with a 5% root-mean-square noise-to-signal ratio.

In discrete-time the system is written, in an recursive form, as:

$$\mathbf{z_k} = \begin{bmatrix} \mathbf{x_{k-1}} + \Delta t \cdot \mathbf{x}_{k-1}^{\cdot} \\ \mathbf{x}_{k-1}^{\cdot} + \Delta t \cdot \ddot{\mathbf{x}}_{k-1} \\ \boldsymbol{\theta}_{k-1} \end{bmatrix}$$

$$(36)$$

and, using Eq. (34) to replace the accelerations $\ddot{\mathbf{x}}_{k-1}$, the process Eq. (11) is written as:

$$\mathbf{z_k} = \begin{bmatrix} z_{1(k-1)} + \Delta t \cdot z_{4(k-1)} \\ z_{2(k-1)} + \Delta t \cdot z_{5(k-1)} \\ z_{3(k-1)} + \Delta t \cdot z_{6(k-1)} \\ z_{4(k-1)} + \Delta t \cdot m_1^{-1} \left\{ u_{1(k-1)}^{e} - (z_{7(k-1)} + z_{8(k-1)})z_{4(k-1)} + z_{8(k-1)}z_{5(k-1)} \\ -(z_{10(k-1)} + z_{11(k-1)})z_{1(k-1)} + z_{11(k-1)}z_{2(k-1)} \right\} \\ z_{5(k-1)} + \Delta t \cdot m_2^{-1} \left\{ u_{2(k-1)}^{e} + z_{8(k-1)}z_{4(k-1)} - (z_{8(k-1)} + z_{9(k-1)})z_{5(k-1)} + z_{9(k-1)}z_{6(k-1)} \\ + z_{11(k-1)}z_{1(k-1)} - (z_{11(k-1)} + z_{12(k-1)})z_{2(k-1)} + z_{12(k-1)}z_{3(k-1)} \right\} \\ z_{6(k-1)} + \Delta t \cdot m_3^{-1} \left\{ u_{3(k-1)}^{e} + z_{9(k-1)}z_{5(k-1)} - z_{9(k-1)}z_{6(k-1)} + z_{12(k-1)}z_{2(k-1)} - z_{12(k-1)}z_{3(k-1)} \right\} \\ z_{7(k-1)} \\ z_{8(k-1)} \\ z_{9(k-1)} \\ z_{10(k-1)} \\ z_{11(k-1)} \\ z_{12(k-1)} \end{bmatrix}$$

$$(37)$$

where,

$$\mathbf{z_k} = \begin{bmatrix} x_{1k} & x_{2k} & x_{3k} & \dot{x}_{1k} & \dot{x}_{2k} & \dot{x}_{3k} & c_{1k} & c_{2k} & c_{3k} & k_{1k} & k_{2k} & k_{3k} \end{bmatrix}^T$$

$$(38)$$

Three initial parameter sets are examined. Set 1 underestimates their value by 50%, Set 2 underestimates their value by 25%, and Set 3 overestimates their value by 50%.

Importantly, acceleration measurements are used for all DOFs, where they are double integrated for displacement and velocity pseudo-measurements. The process covariance $\mathbf{Q_{k-1}}$ and the measurement covariance $\mathbf{R_k}$ matrices are chosen to be constant during the identification process and equal to $10^{-9} \cdot \mathbf{I_{12 \times 12}}$ and $10^{-3} \cdot \mathbf{I_{9 \times 9}}$, respectively. For larger values, the algorithm needs more data and time to converge, or it may even diverge. The prior initial parameter covariance matrix $\mathbf{W_I}$ is set equal to the unit one, namely $\mathbf{I_{6 \times 6}}$.

Regarding the parameter estimation error $E_r(k)$ depended to the time step $k$, it is estimated by the absolute value summation of the error, namely:

$$E_r(k) = \sum_{j}^{j_{max}} \left| \frac{\theta_j(k) - \theta_j^{true}}{\theta_j^{true}} \right|$$

$$(39)$$

where, $\theta_j$ refers to the $j$ parameter estimate, and $\theta_j^{true}$ is the true value of the $j$ parameter. The detailed behavior fluctuation of the parameter estimates is shown in [49]. It is also shown in Section 9 for a 6-DOF system.

Two input loads are examined in Figs. 2 and 3: a pulse input of 100 $N$ amplitude and 0.01 s duration applied at the time instant of 5 s, which is not known beforehand, and an ambient input, namely a white noise-type input, of mean value 0 and variance 4. Here, each figure column refers to each identification set, while the bottom plot shows all Kullback–Leibler divergences.

In Fig. 2, the true and the estimated input, the parameter error, and the DOF 2 true and estimated response are shown for the pulse input. The Kullback–Leibler divergence methodology successfully selects the initial parameter Set 2 which provides the identification with the least estimation error.

Furthermore, in Fig. 3, the application is shown for the ambient input. The Kullback–Leibler divergence methodology also successfully selects the initial parameter Set 2 which provides the identification with the least estimation error.

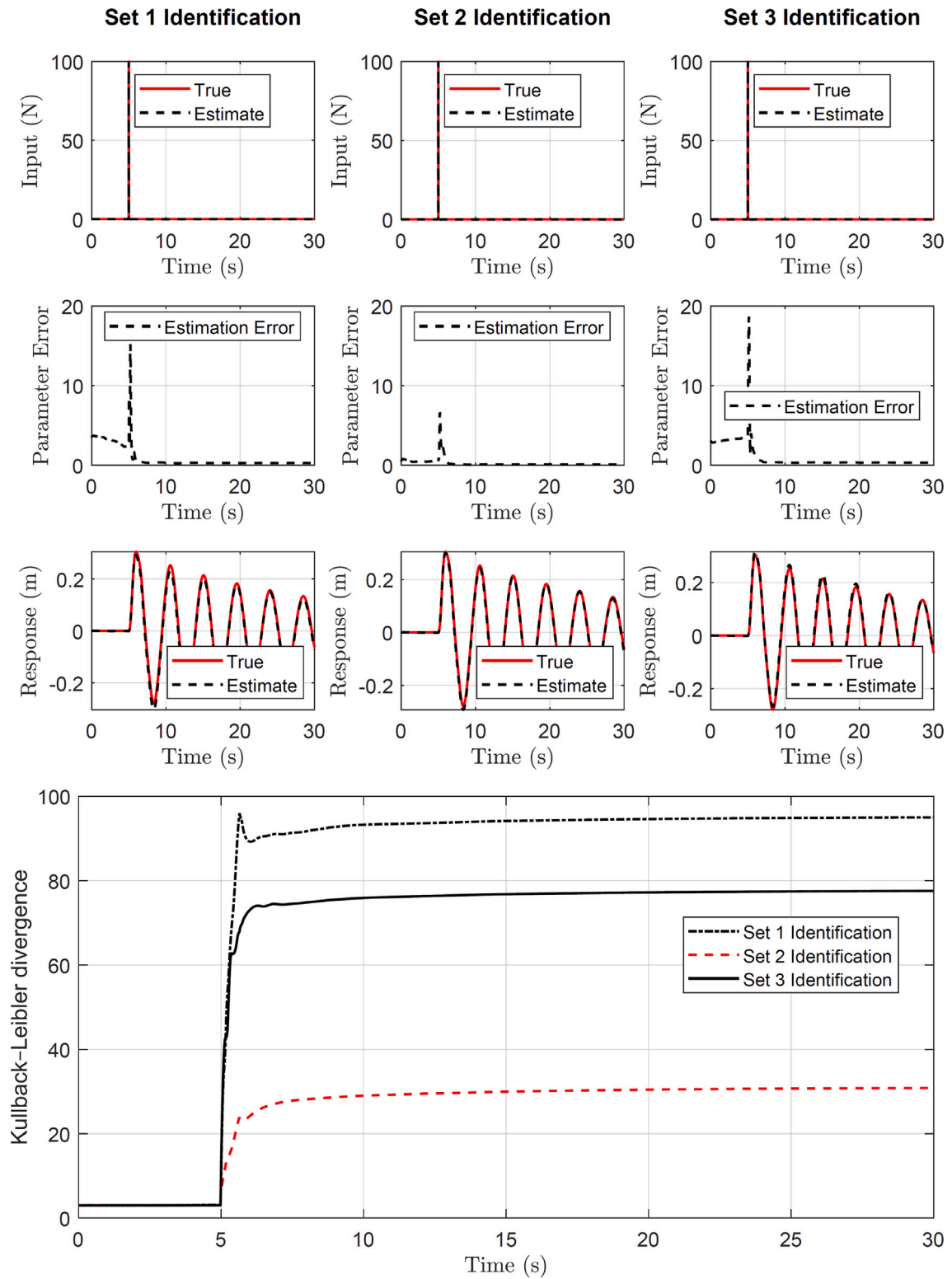

**Fig. 2.** Results for the 3-DOF linear system with pulse input. A detailed discussion on the estimation behavior for each one of the parameters and the input error is provided in Section 9.

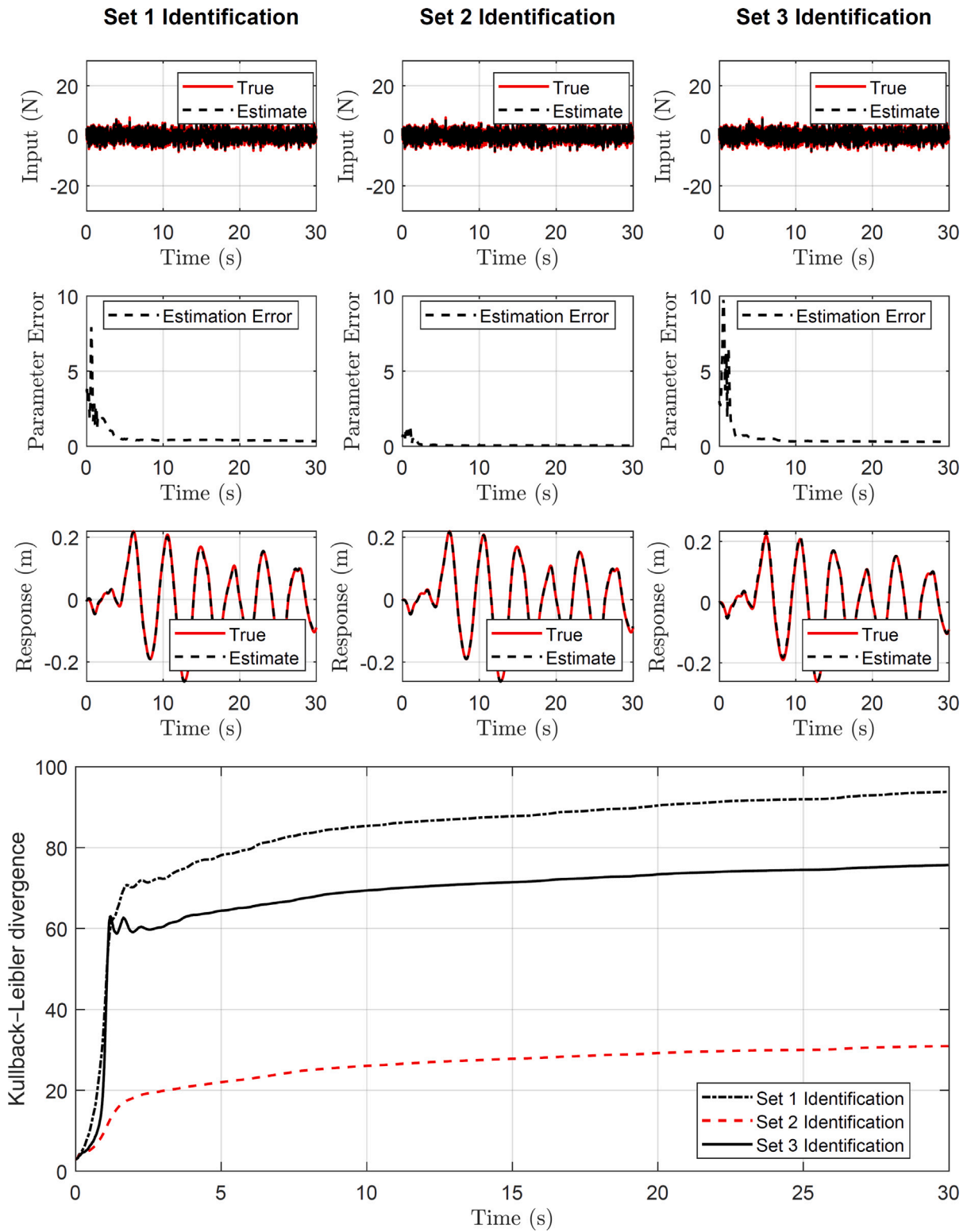

**Fig. 3.** Results for the 3-DOF linear system with white noise input. A detailed discussion on the estimation behavior for each one of the parameters and the input error is provided in Section 9.

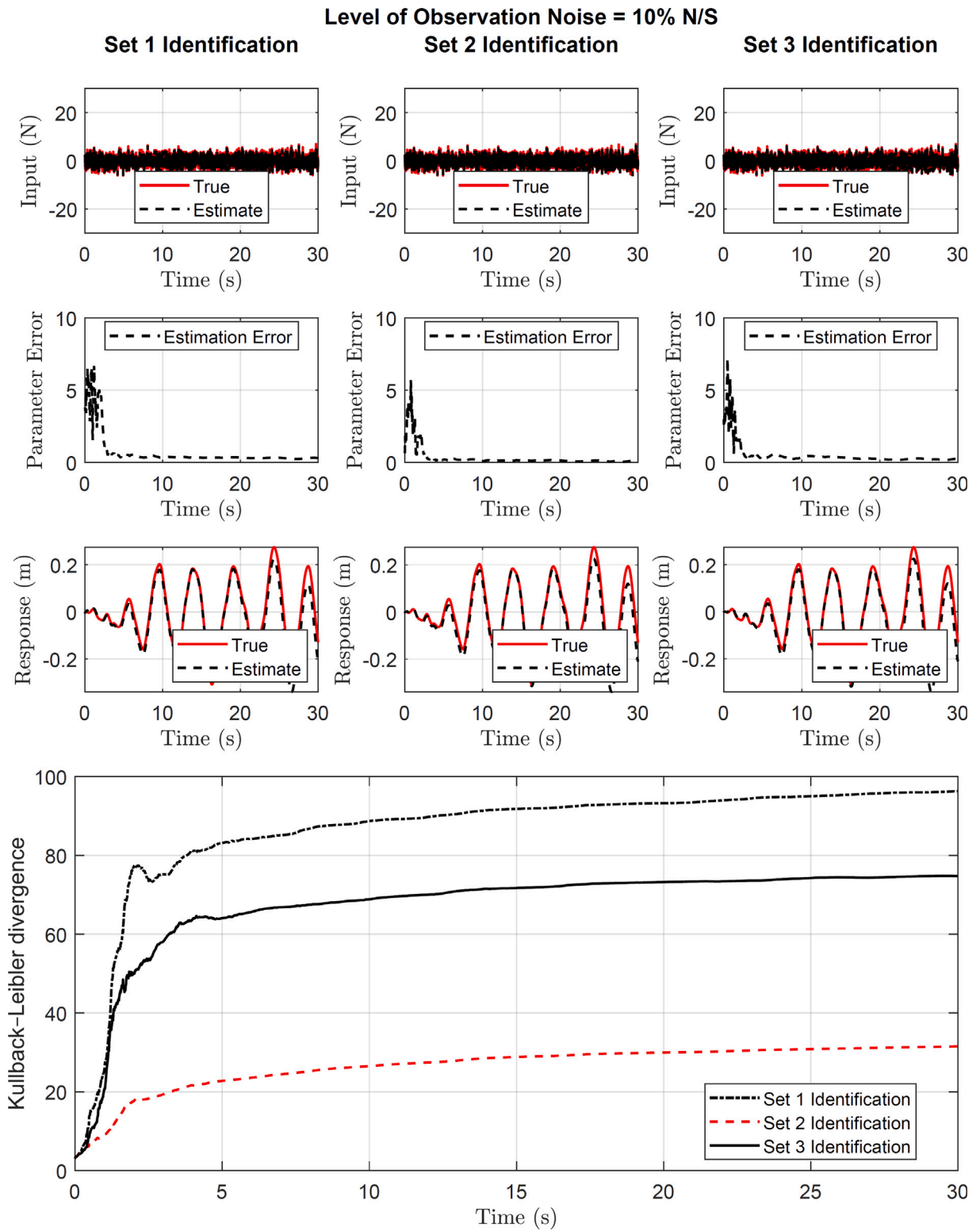

**Fig. 4.** Results for the 3-DOF linear system with white noise input and level of observation noise = 10% root-mean-square noise-to-signal ratio. A detailed discussion on the estimation behavior for each one of the parameters and the input error is provided in Section 9.

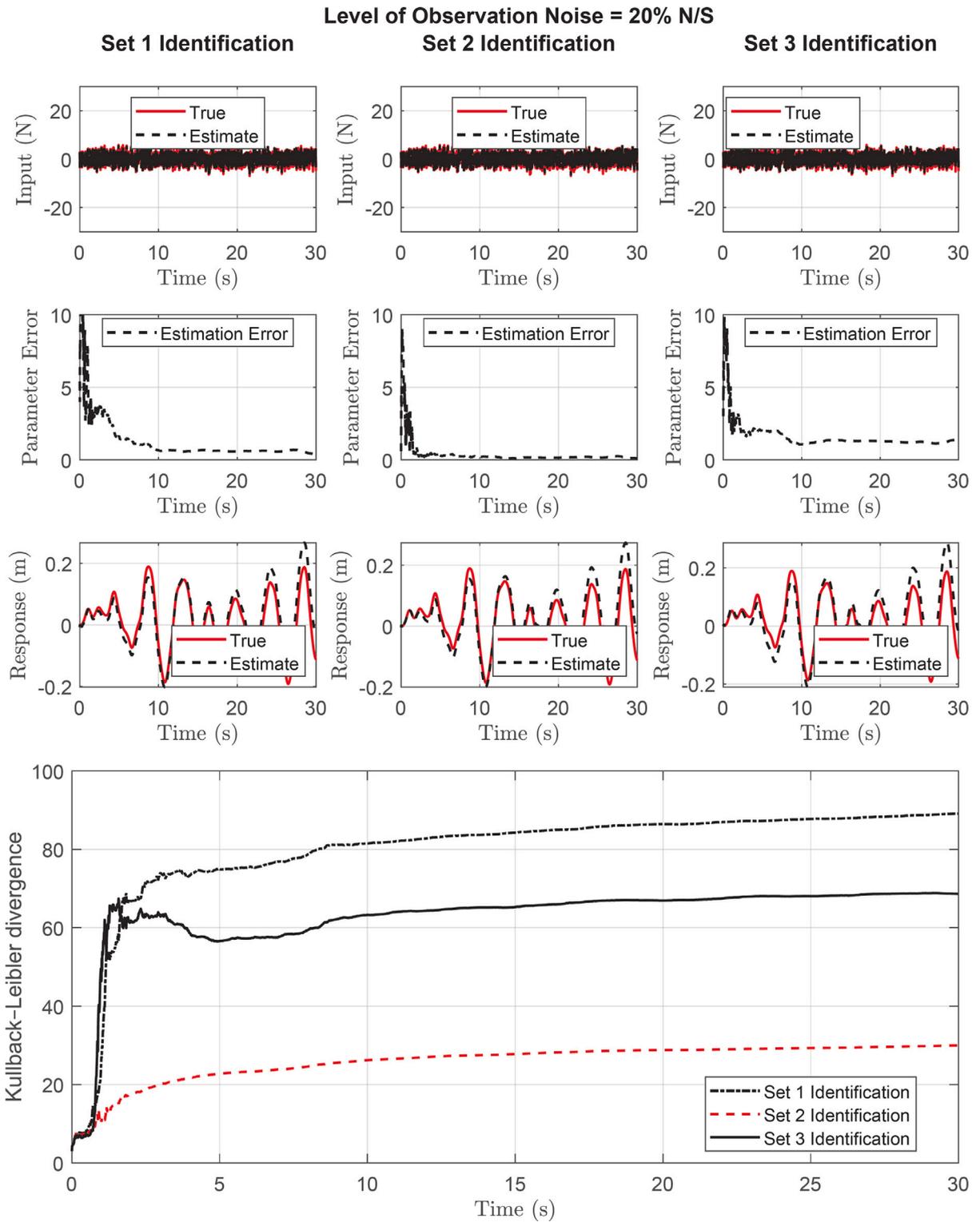

**Fig. 5.** Results for the 3-DOF linear system with white noise input and level of observation noise = 20% root-mean-square noise-to-signal ratio. A detailed discussion on the estimation behavior for each one of the parameters and the input error is provided in Section 9.

Related to the concern of the methodology being tested to various levels of observation noise, Figs. 4 and 5 show additionally the ambient input application when the consider measurement noise for each response signal is contaminated by a Gaussian white noise sequence with a 10% and 20% root-mean-square noise-to-signal ratio, respectively. It is shown that reasonable noise levels can be adequately handed by the methodology. However, for larger noise levels poorer performance is seen on all estimated quantities. Despite that, the better-performed identification indicated by the Kullback–Leibler divergence approach still provides valuable information for the system compared to the less optimal identifications. For all identification sets though in the last noise level case, the state estimation is not satisfactory.

## 7. Application to limited information systems

For the limited information numerical application consider Fig. 1 system of Section 6, with the implementation of the residual Kalman filter with the Kullback–Leibler divergence provided by Table 1B.

The state-space representation is initially written in continuous-time as:

$$\mathbf{z}(t) = \begin{Bmatrix} \mathbf{x}(t) \\ \dot{\mathbf{x}}(t) \end{Bmatrix} \tag{40}$$

The system matrix $\mathbf{A}(\boldsymbol{\theta})$ of $2n \times 2n$ dimension, which contains all the characteristics of the system and depends on the unknown parameters, is written as:

$$\mathbf{A}(\boldsymbol{\theta}) = \begin{bmatrix} \mathbf{0}_{n \times n} & \mathbf{I}_{n \times n} \\ -\mathbf{M}^{-1}\mathbf{K}(\boldsymbol{\theta}) & -\mathbf{M}^{-1}\mathbf{C}(\boldsymbol{\theta}) \end{bmatrix} \tag{41}$$

where, $\mathbf{0}$ and $\mathbf{I}$ are the zero and the unit matrices, respectively. Furthermore, the input distribution matrix $\mathbf{B}$ of $2n \times n$ dimension, which relates the input to the dynamic state vector, is written as:

$$\mathbf{B} = \begin{bmatrix} \mathbf{0}_{n \times n} \\ \mathbf{M}^{-1} \end{bmatrix} \tag{42}$$

The input at the current step is approximated by the system model, namely the equation of motion in the continuous-time at the time instant $(k+1)\,dt$ as:

$$\mathbf{u}_{k+1}^{e} \approx \mathbf{M}_{k+1}^{ma} + [\mathbf{K}_k \quad \mathbf{C}_k]\,\mathbf{z}_{k+1} \tag{43}$$

and then,

$$\mathbf{u}_{k+1}^{e} = \begin{Bmatrix} \mathbf{0}^{known} \text{or} \mathbf{u}_{k+1}^{known} \\ \mathbf{u}_{k+1}^{e} \end{Bmatrix} \tag{44}$$

where, the known zero or non-zero inputs replace the related input rows of the vector $\mathbf{u}_{k+1}^{e}$. The sensitivity matrix at each step is written as:

$$\mathbf{G}_{k+1} = -\left[\frac{\partial\left(\mathbf{M}^{-1}\left(\mathbf{u}_{k+1} - [\mathbf{K} \quad \mathbf{C}]\,\mathbf{z}_{k+1}\right)\right)}{\partial\boldsymbol{\theta}}\right]_{\boldsymbol{\theta} = \boldsymbol{\theta}_{k+1}} \tag{45}$$

which is simplified as:

$$\mathbf{G}_{k+1} = \left[\frac{\partial\left(\mathbf{M}^{-1}\left([\mathbf{K} \quad \mathbf{C}]\,\mathbf{z}_{k+1}\right)\right)}{\partial\boldsymbol{\theta}}\right]_{\boldsymbol{\theta} = \boldsymbol{\theta}_{k+1}} \tag{46}$$

where, the accelerations are replaced by the system model of Eq. (34), and where the input is not a function of the parameter vector in the final form. Importantly, this sensitivity matrix is chosen since the acceleration measurements contain high noise in contrast to the Kalman filter estimated dynamic states (see [50] for further details).

The synthetic measurements are derived in a similar manner to Section 6. The parameter vector is written as:

$$\boldsymbol{\theta} = [k_1, k_2, k_3, c_1, c_2, c_3]^T \tag{47}$$

For accelerometers at DOF 2–3, the observation matrix is written as:

$$\mathbf{H_{23}} = \begin{bmatrix} 0 & \mathbf{1} & 0 & 0 & 0 & 0 \\ 0 & 0 & \mathbf{1} & 0 & 0 & 0 \\ 0 & 0 & 0 & 0 & \mathbf{1} & 0 \\ 0 & 0 & 0 & 0 & 0 & \mathbf{1} \end{bmatrix} \tag{48}$$

mapping the pseudo-measurement vector to the estimated dynamic state vector $\mathbf{z}(k+1)$.

Accordingly, for any number of output (acceleration) measurements, the sensitivity matrix is written as:

$$\mathbf{G}_{k+1} = \mathbf{M}^{-1} \cdot \begin{bmatrix} z_{k+1}(1) & z_{k+1}(1) - z_{k+1}(2) & 0 & z_{k+1}(4) & z_{k+1}(4) - z_{k+1}(5) & 0 \\ 0 & z_{k+1}(2) - z_{k+1}(1) & z_{k+1}(2) - z_{k+1}(3) & 0 & z_{k+1}(5) - z_{k+1}(4) & z_{k+1}(5) - z_{k+1}(6) \\ 0 & 0 & z_{k+1}(3) - z_{k+1}(2) & 0 & 0 & z_{k+1}(6) - z_{k+1}(5) \end{bmatrix} \tag{49}$$

where, the Kalman filtered vectors $\mathbf{z_{k+1}}$ are used.

Three initial parameter sets are examined similarly to Section 6. Importantly, the process covariance $\mathbf{Q_d}$ and the measurement covariance $\mathbf{R_d}$ matrices are chosen to be constant during the identification process and equal to $10^0 \cdot \mathbf{I_{6\times6}}$ and $10^{-10} \cdot \mathbf{I_{6\times6}}$, respectively. The parameter $\lambda^2$ is chosen to be $5 \cdot 10^{-2}$, while the parameter $\mu$ is chosen to be $5 \cdot 10^{-3}$ (see Section 9 for an investigation). The prior initial dynamic state covariance matrix $\mathbf{W_1}$ is set equal to the unit one, namely $\mathbf{I_{6\times6}}$. Regarding the parameter estimation error $E_r(k)$, it is estimated using Eq. (39). The detailed behavior fluctuation of the parameter estimates is shown in [50]. It is also shown in Section 9 for a 6-DOF system.

Two cases are examined on Figs. 6 and 7, where the input is white noise with mean value equal to 0 and variance 4. Input without zero mean value is also examined. Here, each figure column refers to each identification set, while the bottom plot shows all Kullback–Leibler divergences.

In Fig. 6, the true and the estimated input, the parameter error, and the DOF 2 true and estimated response are shown, when the acceleration is measured at DOF 2 and 3. The Kullback–Leibler divergence methodology successfully selects the initial parameter Set 2 which provides the identification with the least estimation error.

Additionally, in Fig. 7, the true and the estimated input, the parameter error, and the DOF 2 true and estimated response are shown, when the acceleration is measured at DOF 2 and 3, and when there is real-time damage. Specifically, the damage is modeled by a parameter change of 50% at the time instant of $50s$. The Kullback–Leibler divergence methodology successfully selects the initial parameter Set 2 which provides the identification with the least estimation error. Interestingly, at the moment of damage, all of the estimated quantities diverge temporarily until they converge to the new ones.

However, for a larger number of DOFs, a larger number of known zero-valued inputs may exist. In that case, a better limited information performance is possible. This is examined using a 6-DOF system.

For this numerical application consider the 6-DOF system described by the following equation:

$$\mathbf{M}\begin{Bmatrix} \ddot{x}_1(t) \\ \ddot{x}_2(t) \\ \ddot{x}_3(t) \\ \ddot{x}_4(t) \\ \ddot{x}_5(t) \\ \ddot{x}_6(t) \end{Bmatrix} + \mathbf{C}\begin{Bmatrix} \dot{x}_1(t) \\ \dot{x}_2(t) \\ \dot{x}_3(t) \\ \dot{x}_4(t) \\ \dot{x}_5(t) \\ \dot{x}_6(t) \end{Bmatrix} + \mathbf{K}\begin{Bmatrix} x_1(t) \\ x_2(t) \\ x_3(t) \\ x_4(t) \\ x_5(t) \\ x_6(t) \end{Bmatrix} = \begin{Bmatrix} 0 \\ 0 \\ 0 \\ 0 \\ u_5(t) \\ u_6(t) \end{Bmatrix} \tag{50}$$

for zero or non-zero input $u_5(t)$ at DOF 5, where the system matrices which need to be identified (apart from $\mathbf{M}$) are:

$$\mathbf{M} = \begin{bmatrix} m_1 & 0 & 0 & 0 & 0 & 0 \\ 0 & m_2 & 0 & 0 & 0 & 0 \\ 0 & 0 & m_3 & 0 & 0 & 0 \\ 0 & 0 & 0 & m_4 & 0 & 0 \\ 0 & 0 & 0 & 0 & m_5 & 0 \\ 0 & 0 & 0 & 0 & 0 & m_6 \end{bmatrix} = \begin{bmatrix} 1 & 0 & 0 & 0 & 0 & 0 \\ 0 & 1 & 0 & 0 & 0 & 0 \\ 0 & 0 & 1 & 0 & 0 & 0 \\ 0 & 0 & 0 & 1 & 0 & 0 \\ 0 & 0 & 0 & 0 & 1 & 0 \\ 0 & 0 & 0 & 0 & 0 & 1 \end{bmatrix},$$

$$\mathbf{C} = \begin{bmatrix} c_1+c_2 & -c_2 & 0 & 0 & 0 & 0 \\ -c_2 & c_2+c_3 & -c_3 & 0 & 0 & 0 \\ 0 & -c_3 & c_3+c_4 & -c_4 & 0 & 0 \\ 0 & 0 & -c_4 & c_4+c_5 & -c_5 & 0 \\ 0 & 0 & 0 & -c_5 & c_5+c_6 & -c_6 \\ 0 & 0 & 0 & 0 & -c_5 & c_6 \end{bmatrix} = \begin{bmatrix} 0.25+0.25 & -0.25 & 0 & 0 & 0 & 0 \\ -0.25 & 0.25+0.5 & -0.5 & 0 & 0 & 0 \\ 0 & -0.5 & 0.5+0.5 & -0.5 & 0 & 0 \\ 0 & 0 & -0.5 & 0.5+0.75 & -0.75 & 0 \\ 0 & 0 & 0 & -0.75 & 0.75+0.75 & -0.75 \\ 0 & 0 & 0 & 0 & -0.75 & 0.75 \end{bmatrix},$$

$$\mathbf{K} = \begin{bmatrix} k_1+k_2 & -k_2 & 0 & 0 & 0 & 0 \\ -k_2 & k_2+k_3 & -k_3 & 0 & 0 & 0 \\ 0 & -k_3 & k_3+k_4 & -k_4 & 0 & 0 \\ 0 & 0 & -k_4 & k_4+k_5 & -k_5 & 0 \\ 0 & 0 & 0 & -k_5 & k_5+k_6 & -k_6 \\ 0 & 0 & 0 & 0 & -k_5 & k_6 \end{bmatrix} = \begin{bmatrix} 9+9 & -9 & 0 & 0 & 0 & 0 \\ -9 & 9+11 & -11 & 0 & 0 & 0 \\ 0 & -11 & 11+11 & -11 & 0 & 0 \\ 0 & 0 & -11 & 11+13 & -13 & 0 \\ 0 & 0 & 0 & -13 & 13+13 & -13 \\ 0 & 0 & 0 & 0 & -13 & 13 \end{bmatrix}$$

with initial conditions $\mathbf{x}(0) = [0 \quad 0 \quad 0 \quad 0 \quad 0 \quad 0]^T$ and $\dot{\mathbf{x}}(0) = [0 \quad 0 \quad 0 \quad 0 \quad 0 \quad 0]^T$.

The synthetic measurements are derived in a similar manner to Section 6. The parameter vector is written as:

$$\boldsymbol{\theta} = [k_1, k_2, k_3, k_4, k_5, k_6, c_1, c_2, c_3 c_4, c_5, c_6]^T \tag{51}$$

The observation matrices are written in a similar manner to the previous system. Here, the acceleration sensors which are removed each time, are chosen to be as far as possible from the DOF where the input is applied, which is a crucial detail for the success of the input–parameter–state estimation.

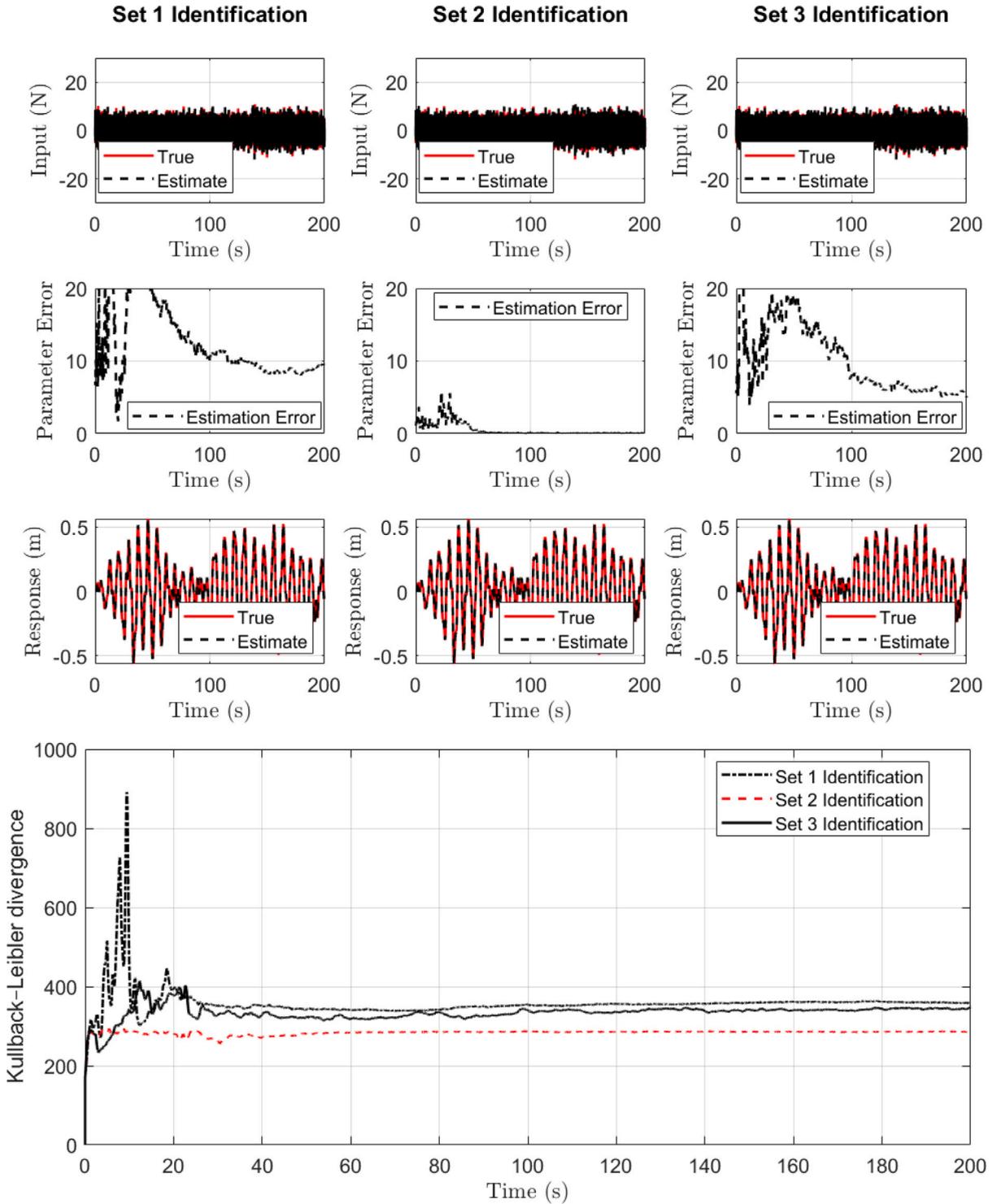

**Fig. 6.** Results for the 3-DOF linear system when measuring the acceleration at DOF 2 to 3. A detailed discussion on the estimation behavior for each one of the parameters and the input error is provided in Section 9.

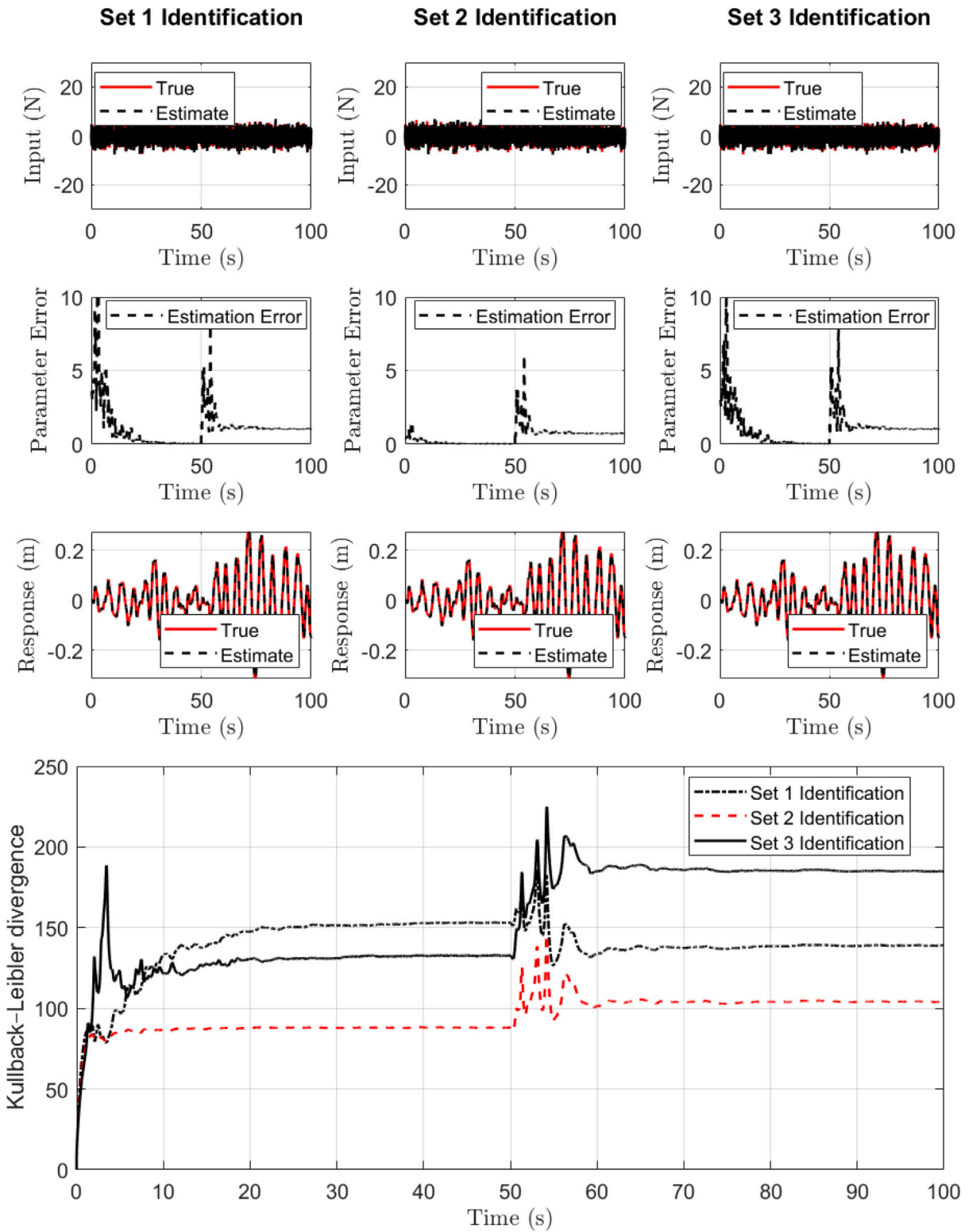

**Fig. 7.** Results for the 3-DOF linear system with 50% damage considered for the system parameters at the time instant 50 s, when measuring the acceleration at DOF 2 to 3. A detailed discussion on the estimation behavior for each one of the parameters and the input error is provided in Section 9.

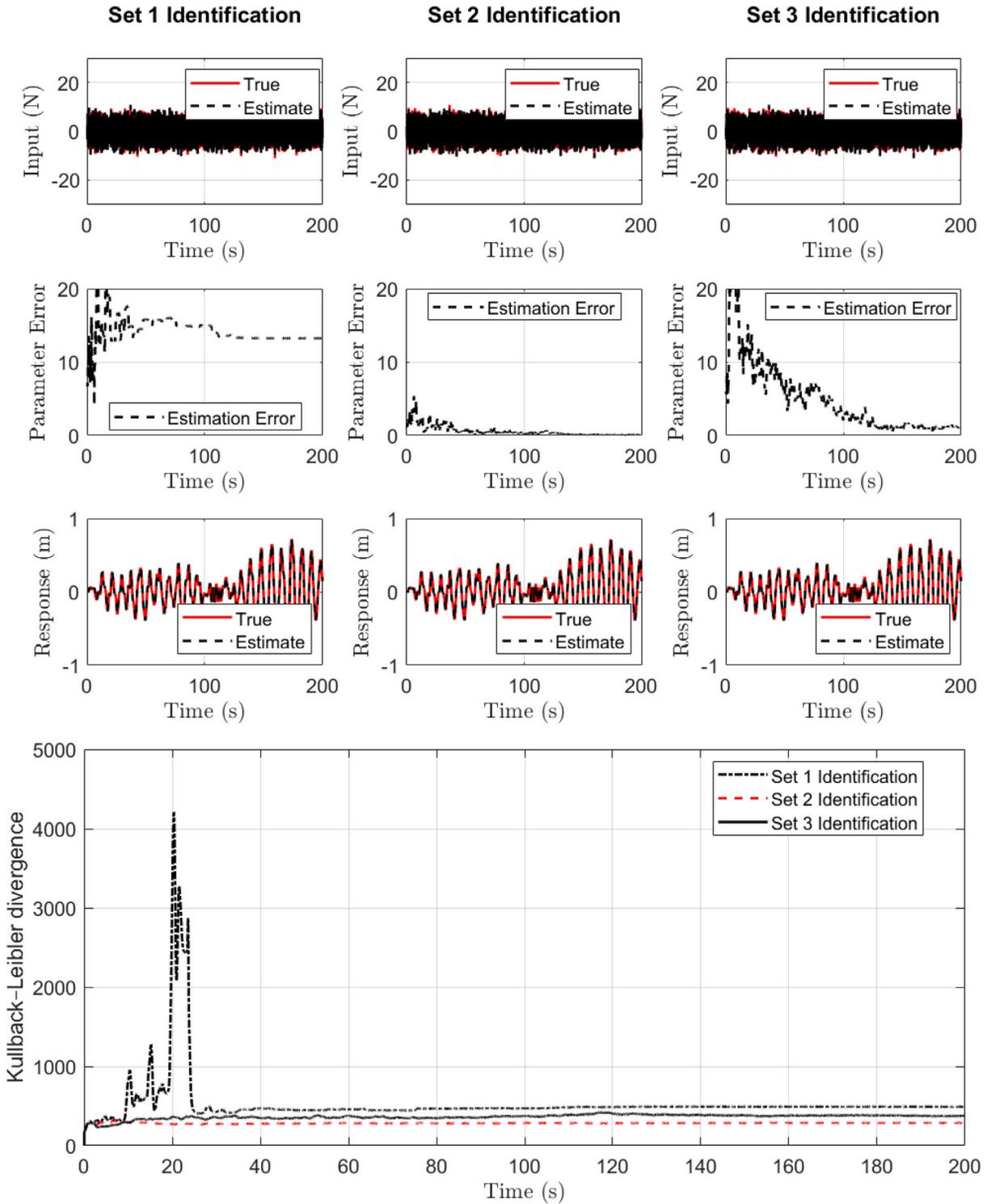

**Fig. 8.** Results for the 6-DOF linear system when measuring the acceleration at DOF 3 to 6. A detailed discussion on the estimation behavior for each one of the parameters and the input error is provided in Section 9.

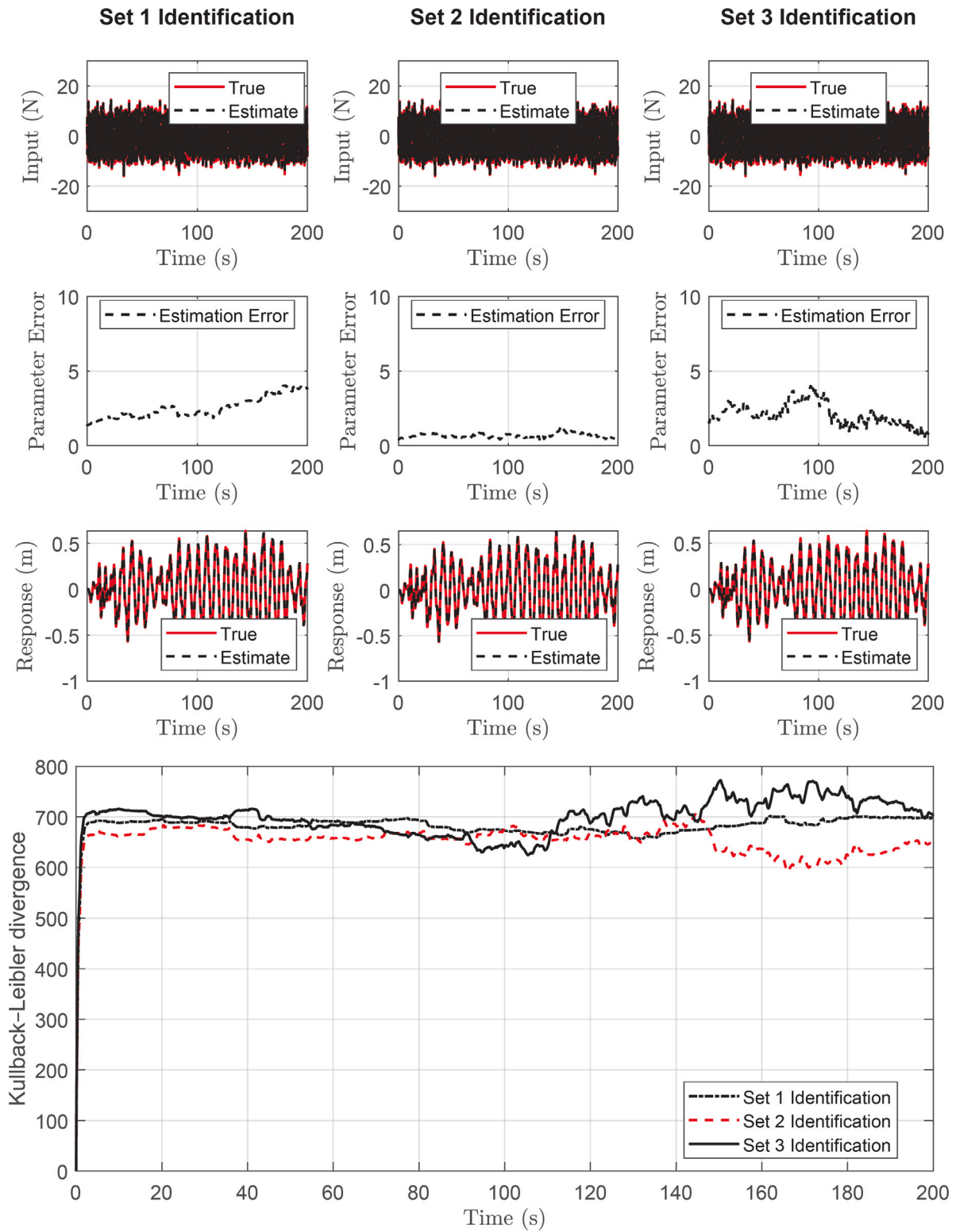

**Fig. 9.** Results for the 6-DOF linear system when measuring the acceleration at DOF 4 to 6. A detailed discussion on the estimation behavior for each one of the parameters and the input error is provided in Section 9.

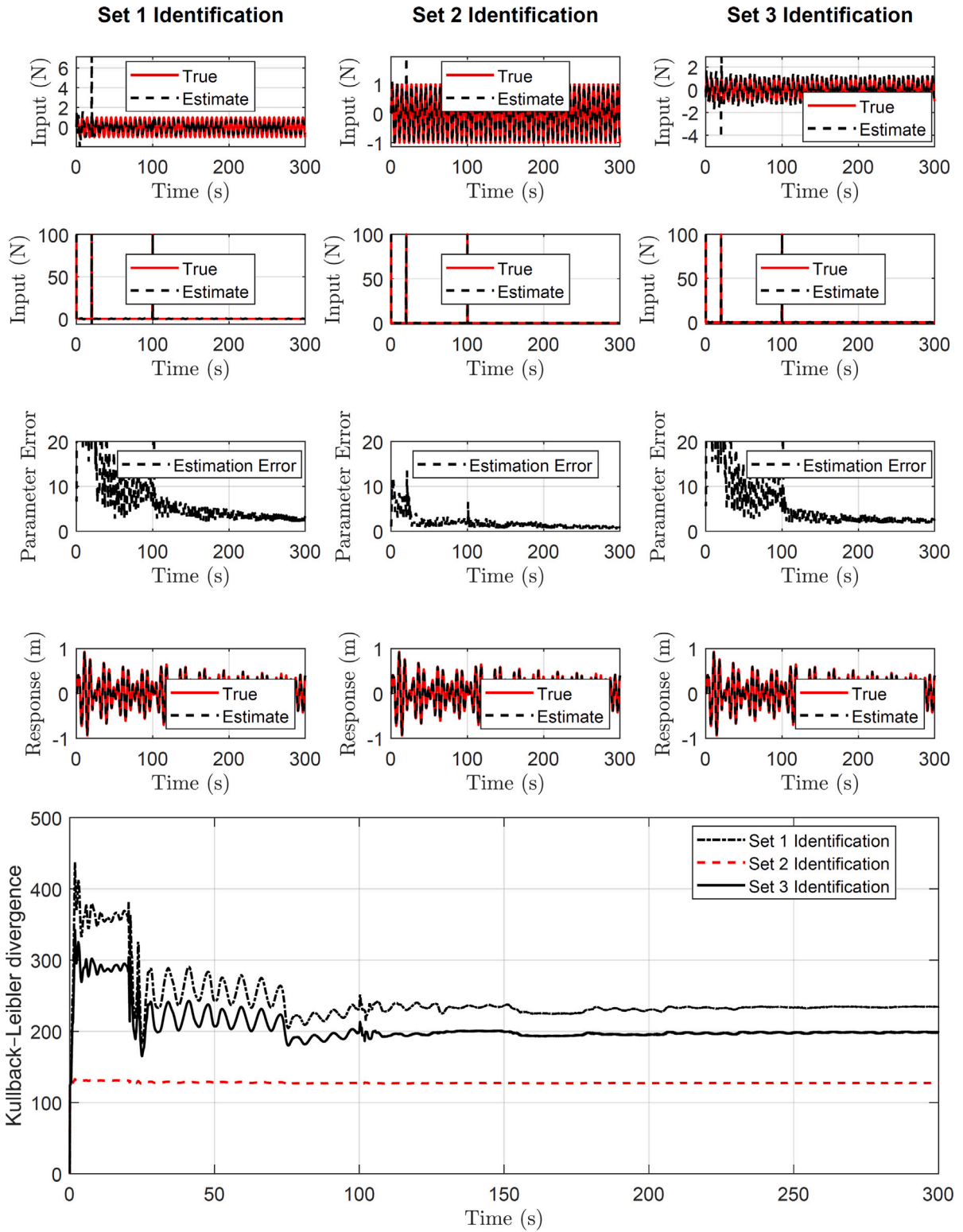

**Fig. 10.** Results for the 6-DOF linear system with two non-white inputs when measuring the acceleration at DOF 4 to 6. A detailed discussion on the estimation behavior for each one of the parameters and the input error is provided in Section 9.

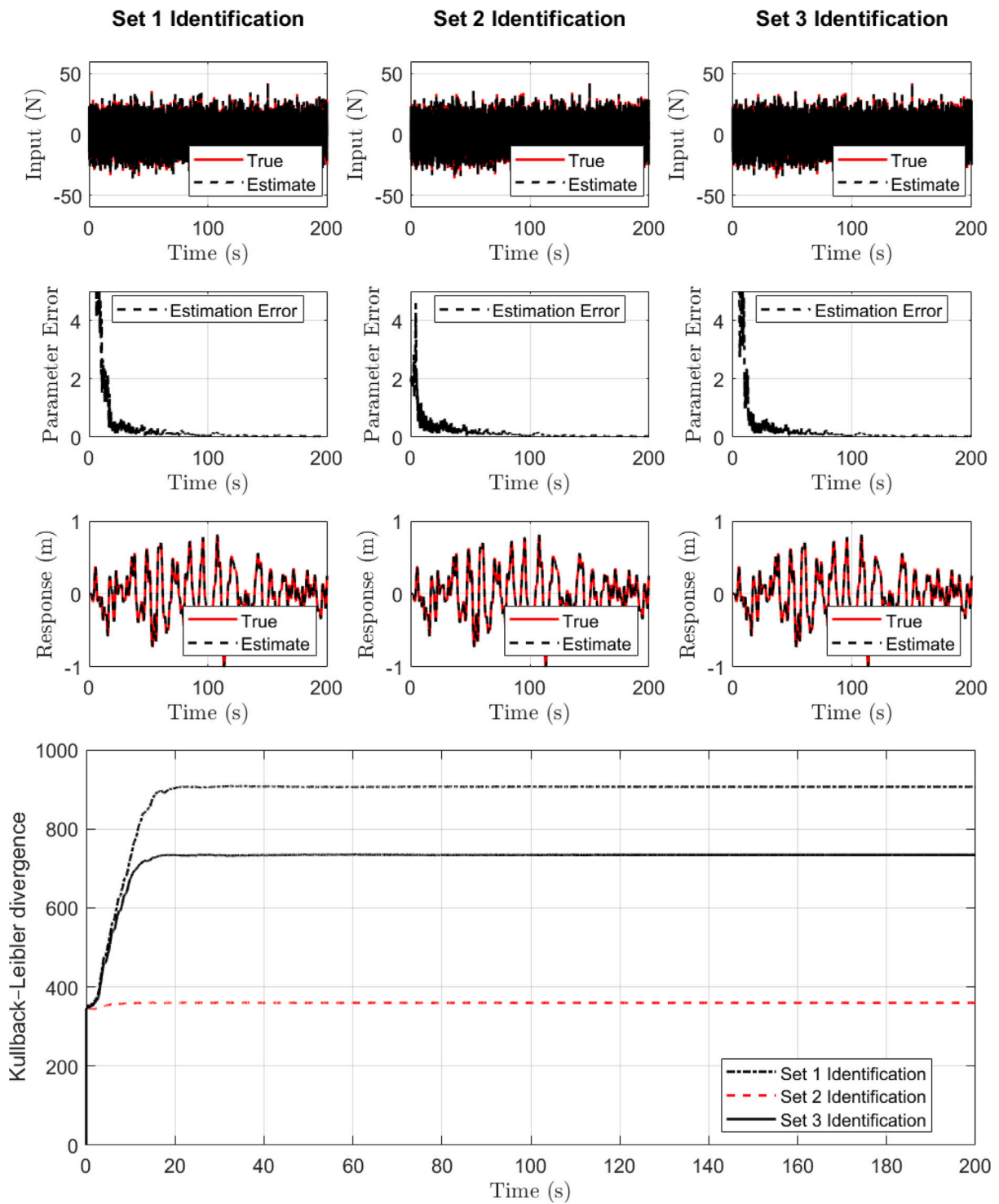

**Fig. 11.** Results for a 10-DOF linear system when measuring the acceleration at DOF 5 to 10. A detailed discussion on the estimation behavior for each one of the parameters and the input error is provided in Section 9.

The sensitivity matrix for this system, for any number of accelerometers and inputs, is written as:

$$\mathbf{G_{k+1}} = \mathbf{M}^{-1}$$

$$
\begin{bmatrix}
z(1) & z(1)-z(2) & 0 & 0 & 0 & 0 & z(7) & z(7)-z(8) & 0 & 0 & 0 & 0 \\
0 & z(2)-z(1) & z(2)-z(3) & 0 & 0 & 0 & 0 & z(8)-z(7) & z(8)-z(9) & 0 & 0 & 0 \\
0 & 0 & z(3)-z(2) & z(3)-z(4) & 0 & 0 & 0 & 0 & z(9)-z(8) & z(9)-z(10) & 0 & 0 \\
0 & 0 & 0 & z(4)-z(3) & z(4)-z(5) & 0 & 0 & 0 & 0 & z(10)-z(9) & z(10)-z(11) & 0 \\
0 & 0 & 0 & 0 & z(5)-z(4) & z(5)-z(6) & 0 & 0 & 0 & 0 & z(11)-z(10) & z(11)-z(12) \\
0 & 0 & 0 & 0 & 0 & z(6)-z(5) & 0 & 0 & 0 & 0 & 0 & z(12)-z(11)
\end{bmatrix}
$$

$$(52)$$

where, the Kalman filtered $\mathbf{z}$ vector refers to the $\mathbf{z_{k+1}}$ estimates, and the parenthesis value denotes the vector row.

Three initial parameter sets are examined similarly to Section 6. Importantly, the process covariance $\mathbf{Q_d}$ and the measurement covariance $\mathbf{R_d}$ matrices are chosen to be constant during the identification process and equal to $10^0 \cdot \mathbf{I_{12 \times 12}}$ and $10^{-10} \cdot \mathbf{I_{12 \times 12}}$, respectively. The parameter $\lambda^2$ is chosen to be $5 \cdot 10^{-2}$, while the parameter $\mu$ is chosen to be $5 \cdot 10^{-3}$ (see Section 9 for an investigation). The prior initial dynamic state covariance matrix is set equal to the unit one, namely $\mathbf{I_{12 \times 12}}$. Regarding the parameter estimation error $E_r(k)$, it is estimated using Eq. (39). The detailed behavior fluctuation of the parameter estimates is shown in [50]. It is also shown in Section 9 for a 6-DOF system.

Three cases are examined on Figs. 8–10. Here, each figure column refers to each identification set, while the bottom plot shows all Kullback–Leibler divergences.

In Fig. 8, the true and the estimated input, the parameter error, and the DOF 2 true and estimated response are shown for a white noise input with mean value equal to 0 and variance 9, when acceleration are measured at DOFs 3 to 6. The Kullback–Leibler divergence methodology successfully selects the initial parameter Set 2 which provides the identification with the least estimation error.

In Fig. 9, the true and the estimated input, the parameter error, and the DOF 2 true and estimated response are shown for a white noise input with mean value equal to 0 and variance 9, when acceleration are measured at DOFs 4 to 6. The Kullback–Leibler divergence methodology successfully selects the initial parameter Set 2 which provides the identification with the least estimation error.

Finally, in Fig. 10, the true and the estimated input, the parameter error, and the DOF 2 true and estimated response are shown for two input loads, when acceleration are measured at DOFs 4 to 6. Specifically, two simultaneous non-white noise inputs are examined; three pulses of $100\,\mathrm{N}$ for $0.01\mathrm{s}$ are applied at DOF 6 at random and unknown time instants, and a harmonic load of amplitude 1.0 N and circular frequency of 1.0 rad/s is applied at DOF 5. The Kullback–Leibler divergence methodology successfully selects the initial parameter Set 2 which provides the identification with the least estimation error. Interestingly here, Set 1 and Set 3 identifications have a considerable input estimation error at DOF 5.

Related to the concern of the methodology being tested to structures with even larger DOFs, it is shown that an even larger number of DOFs provides more information for the identification as there are potential more DOFs with a known zero-valued input. This results in a better identification performance with a much faster convergence. Importantly, all the models used for identification do not include any parametrization. This makes the identification more complex compared to even large finite element models where very few parameters are often identified such as some Young's moduli, yield stresses, and one or two Rayleigh damping parameters. Fig. 11 shows the true and the estimated input, the parameter error, and the DOF 10 true and estimated response for a 10-DOF system with 20 stiffness and damping unknown parameters, when acceleration are measured at DOFs 5 to 10. The Kullback–Leibler divergence methodology successfully selects the initial parameter Set 2 which provides the identification with the least estimation error.

## 8. Application to nonlinear systems

For the nonlinear numerical application consider the Duffing nonlinear 2-DOF system with the implementation of the unscented Kalman filter with the Kullback–Leibler divergence provided by Table 1A. The system is described by the following equation:

$$
\mathbf{M}\begin{Bmatrix} \ddot{x}_1(t) \\ \ddot{x}_2(t) \end{Bmatrix} + \mathbf{C}\begin{Bmatrix} \dot{x}_1(t) \\ \dot{x}_2(t) \end{Bmatrix} + \mathbf{K}\begin{Bmatrix} x_1(t) \\ x_2(t) \end{Bmatrix} + \mathbf{E}\begin{Bmatrix} x_1^3(t) \\ (x_2(t)-x_1(t))^3 \end{Bmatrix} = \begin{Bmatrix} 0 \\ u_2(t) \end{Bmatrix}
$$

$$(53)$$

where the system matrices which need to be identified (apart from $\mathbf{M}$) are:

$$
\mathbf{M} = \begin{bmatrix} m_1 & 0 \\ 0 & m_2 \end{bmatrix} = \begin{bmatrix} 1 & 0 \\ 0 & 1 \end{bmatrix}, \quad
\mathbf{C} = \begin{bmatrix} c_1+c_2 & -c_2 \\ -c_2 & c_2 \end{bmatrix} = \begin{bmatrix} 0.5+0.5 & -0.5 \\ -0.5 & 0.5 \end{bmatrix},
$$
$$
\mathbf{K} = \begin{bmatrix} k_1+k_2 & -k_2 \\ -k_2 & k_2 \end{bmatrix} = \begin{bmatrix} 3+4.5 & -4.5 \\ -4.5 & 4.5 \end{bmatrix}, \quad
\mathbf{E} = \begin{bmatrix} \epsilon_1 & -\epsilon_2 \\ 0 & \epsilon_2 \end{bmatrix} = \begin{bmatrix} 15 & -27 \\ 0 & 27 \end{bmatrix}
$$

$$(54)$$

with initial conditions $\mathbf{x}(0) = [0 \quad 0]^T$ and $\dot{\mathbf{x}}(0) = [0 \quad 0]^T$. A pulse input of $100\ N$ for $0.01$ s is applied at the time instant of 5 s, which is not known beforehand. Simultaneously, a white noise-type input of mean value 0 and variance 4 is applied to the system. Both excitations are applied at DOF 2. The synthetic measurements are created in a similar manner to Section 6.

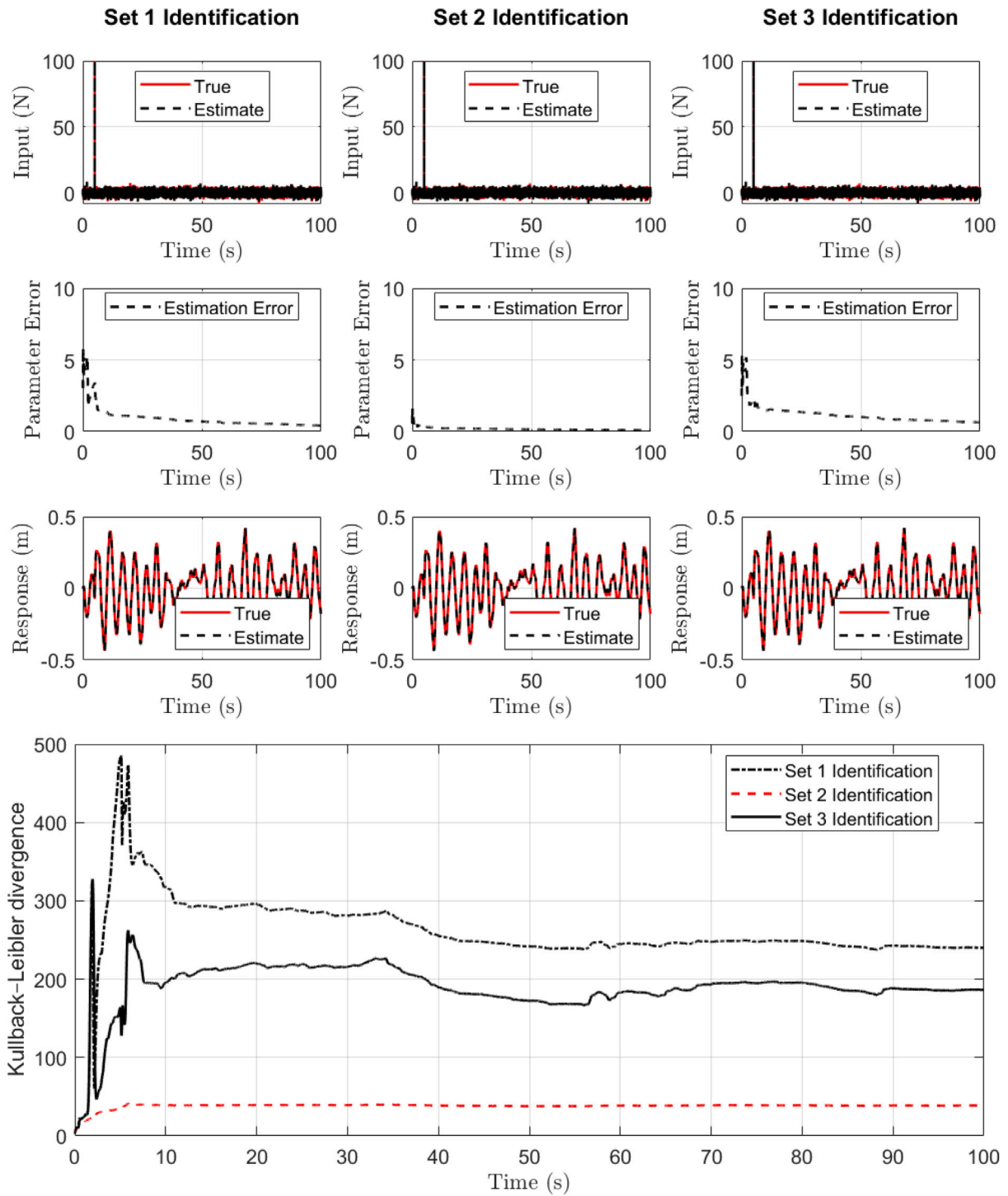

**Fig. 12.** Results for the 2-DOF nonlinear system with pulse and white noise input when measuring the acceleration and displacement at DOF 1 to 2. A detailed discussion on the estimation behavior for each one of the parameters and the input error is provided in Section 9.

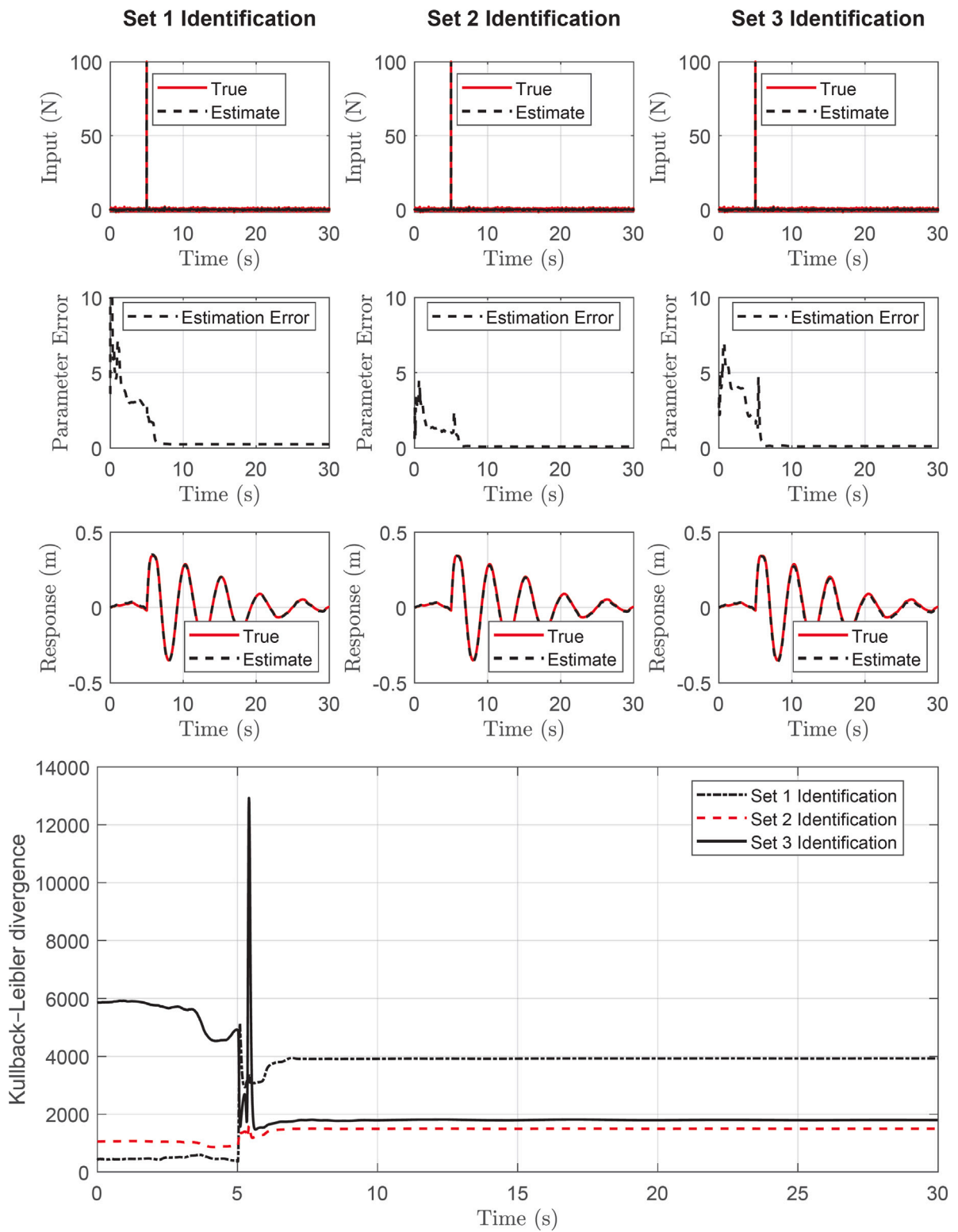

**Fig. 13.** Results for the 2-DOF nonlinear system with pulse and white noise input when measuring the acceleration and displacement at DOF 2. A detailed discussion on the estimation behavior for each one of the parameters and the input error is provided in Section 9.

In discrete time the system can be written, in an recursive form, as:

$$\mathbf{z_k} = \begin{bmatrix} \mathbf{x_{k-1}} + \Delta t \cdot \mathbf{x}_{k-1}^{\cdot} \\ \mathbf{x}_{k-1}^{\cdot} + \Delta t \cdot \ddot{\mathbf{x}}_{k-1} \\ \boldsymbol{\theta_{k-1}} \end{bmatrix} \tag{55}$$

and, using the equation of motion (34) to replace the accelerations $\ddot{\mathbf{x}}_{k-1}$, the process Eq. (11) is written as:

$$\mathbf{z_k} = \begin{bmatrix} z_{1(k-1)} + \Delta t \cdot z_{3(k-1)} \\ z_{2(k-1)} + \Delta t \cdot z_{4(k-1)} \\ z_{3(k-1)} + \Delta t \cdot m_1^{-1} \left\{ \begin{array}{l} u_{1(k-1)}^{e\nearrow 0} - (z_{5(k-1)} + z_{6(k-1)})z_{3(k-1)} - (z_{7(k-1)} + z_{8(k-1)})z_{1(k-1)} \\ + z_{6(k-1)}z_{4(k-1)} + z_{8(k-1)}z_{2(k-1)} - z_{9(k-1)}z_{1(k-1)}^3 + z_{10(k-1)}(z_{2(k-1)} - z_{1(k-1)})^3 \end{array} \right\} \\ z_{4(k-1)} + \Delta t \cdot m_2^{-1} \left\{ \begin{array}{l} u_{2(k-1)}^e + z_{6(k-1)}z_{3(k-1)} - z_{6(k-1)}z_{4(k-1)} + z_{8(k-1)}z_{1(k-1)} \\ - z_{8(k-1)}z_{2(k-1)} - z_{10(k-1)}(z_{2(k-1)} - z_{1(k-1)})^3 \end{array} \right\} \\ z_{5(k-1)} \\ z_{6(k-1)} \\ z_{7(k-1)} \\ z_{8(k-1)} \\ z_{9(k-1)} \\ z_{10(k-1)} \end{bmatrix} \tag{56}$$

where,

$$\mathbf{z_k} = \begin{bmatrix} x_{1k} & x_{2k} & \dot{x}_{1k} & \dot{x}_{2k} & c_{1k} & c_{2k} & k_{1k} & k_{2k} & \epsilon_{1k} & \epsilon_{2k} \end{bmatrix}^T \tag{57}$$

Three initial parameter sets are examined similarly to Section 6. Importantly, the process covariance $\mathbf{Q_{k-1}}$ and the measurement covariance $\mathbf{R_k}$ matrices are chosen to be constant during the identification process and equal to $10^{-9} \cdot \mathbf{I_{10 \times 10}}$ and $10^{-5} \cdot \mathbf{I_{6 \times 6}}$, respectively. The prior initial parameter covariance matrix $\mathbf{W_1}$ is set equal to the unit one, namely $\mathbf{I_{6 \times 6}}$. Regarding the parameter estimation error $E_r(k)$, it is estimated using Eq. (39). The detailed behavior fluctuation of the parameter estimates is shown in [49]. It is also shown in Section 9 for a 6-DOF system.

Two cases are examined on Figs. 12 and 13. Here, each figure column refers to each identification set, while the bottom plot shows all Kullback–Leibler divergences.

In Fig. 12, the true and the estimated input, the parameter error, and the DOF 2 true and estimated response are shown. The Kullback–Leibler divergence methodology successfully selects the initial parameter Set 2 which provides the identification with the least estimation error.

Additionally, in Fig. 13 the application is shown when measuring only DOF 2, and the approximation is used for the prior step acceleration estimation with additional displacement sensing. The Kullback–Leibler divergence methodology also successfully selects the initial parameter Set 2 which provides the identification with the least estimation error.

## 9. Discussion

The correct identification of a system when the dynamic states, parameters, and input are unknown requires certain assumptions. Without those, there are multiple erroneous combinations of parameters and inputs which simultaneously satisfy the measured output [57]. As a consequence, regardless of which identification methodology is used, the estimation fails. In the case where the input is known and there is no measurement noise in a linear system, the identifiability of the parameter vector can be checked using the local identifiability test [58]. For nonlinear systems, the local identifiability can be examined using the Observability Rank Condition [59–62].

In the case of the unknown input, a successful input–parameter–state estimation is possible when one of the following takes place:

(a) All inputs are incorporated in the process error. This occurs when the input mean value is equal to zero and the amplitude is low. In such a case, the error derived from the parameters and the dynamic states is also incorporated into the process error. As a result, the update process filters out this pseudo-noise type of input.

(b) One or more inputs, either zero or non-zero, are known. Here, the known input, regardless of whether it is zero or non-zero, can potentially lead to the correct identification of the real parameters, the dynamic states, and as a result the real input at the other DOFs because the algorithm is calibrated based on this.

So far, figures have been provided for the numerical examples to demonstrate the application range of the proposed method. They all consist of four rows, including the input estimation, the parameter error, the response, and the Kullback–Leibler divergence. The vertical axis of the first row plots, though, do not show the input estimation error to provide emphasis to the Kullback–Leibler divergence since the detailed input and parameter estimation results are already available [49,50]. However, this work is presented as a coupled input–parameter–state estimation method, and the information showing the convergence trajectory of different model parameters should be provided. Importantly, the suggested approach aims to choose the set of initial parameters that could yield accurate estimates for the unknown quantities and at this point, an observation at the parameter error plots may show error in estimating model parameters even with good starting points given that all of them are visualized with only one plot.

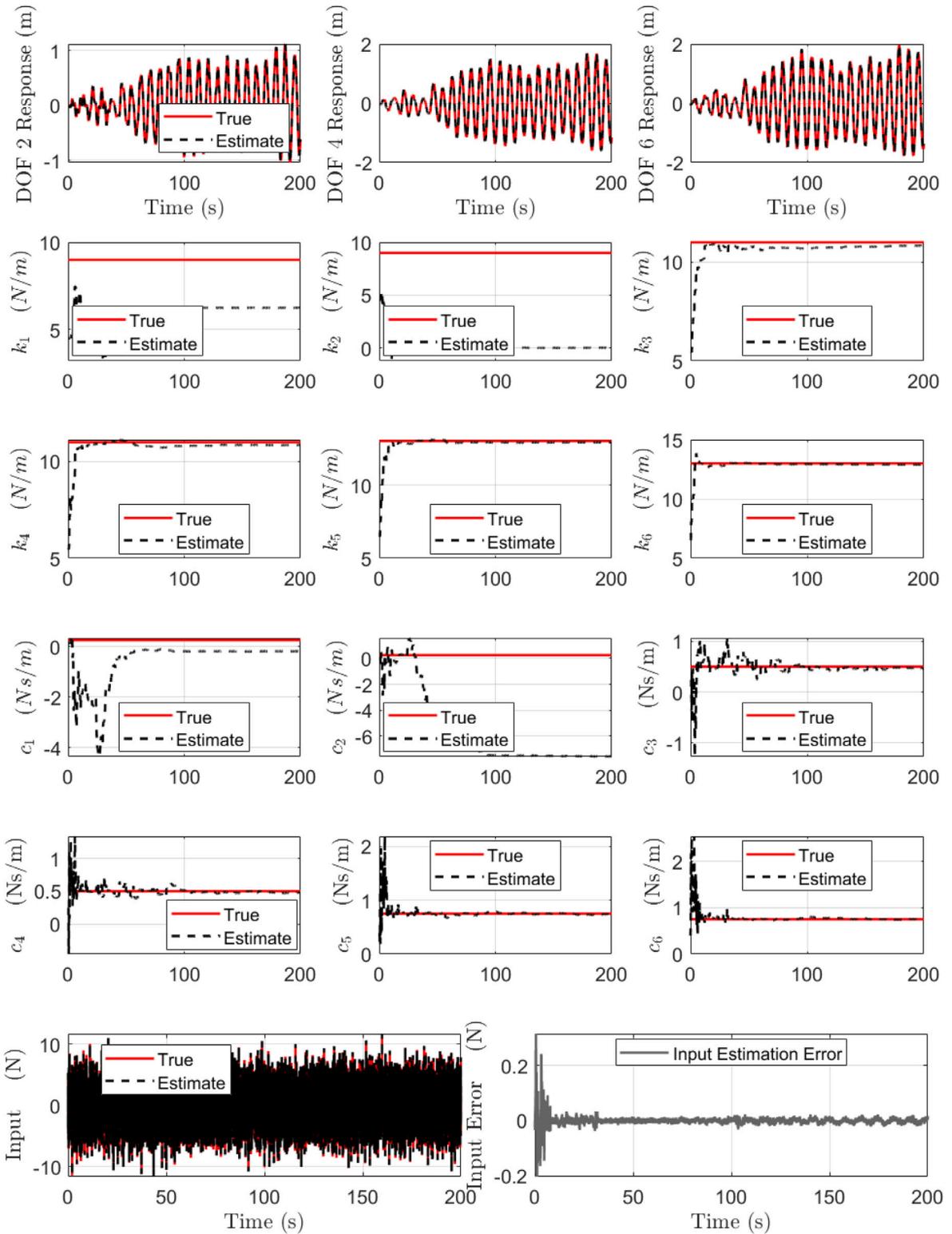

**Fig. 14.** Detailed parameter estimation and input error results for the system of Fig. 8 when the initial parameter set underestimate their value by 50%.

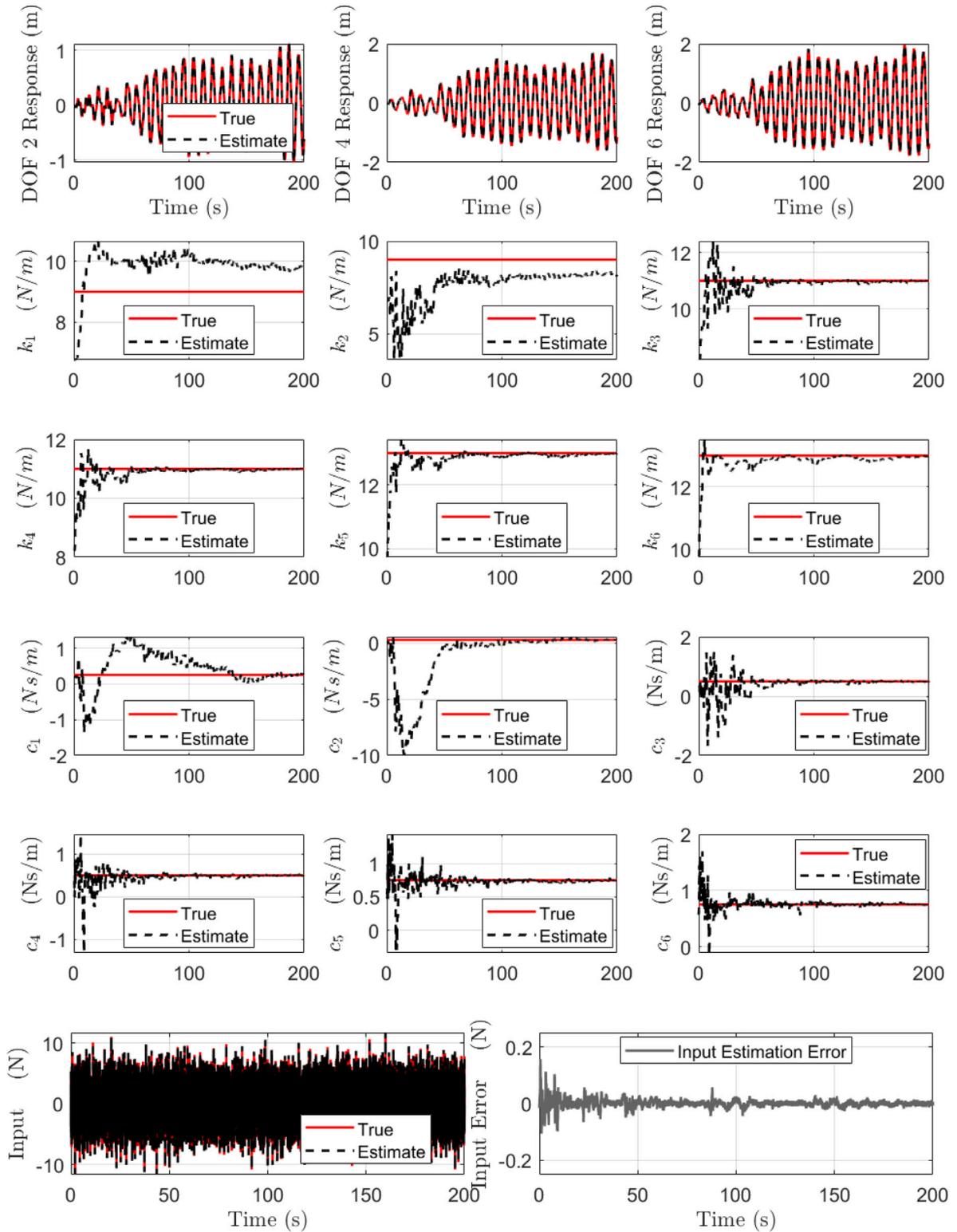

**Fig. 15.** Detailed parameter estimation and input error results for the system of Fig. 8 when the initial parameter set underestimate their value by 25%.

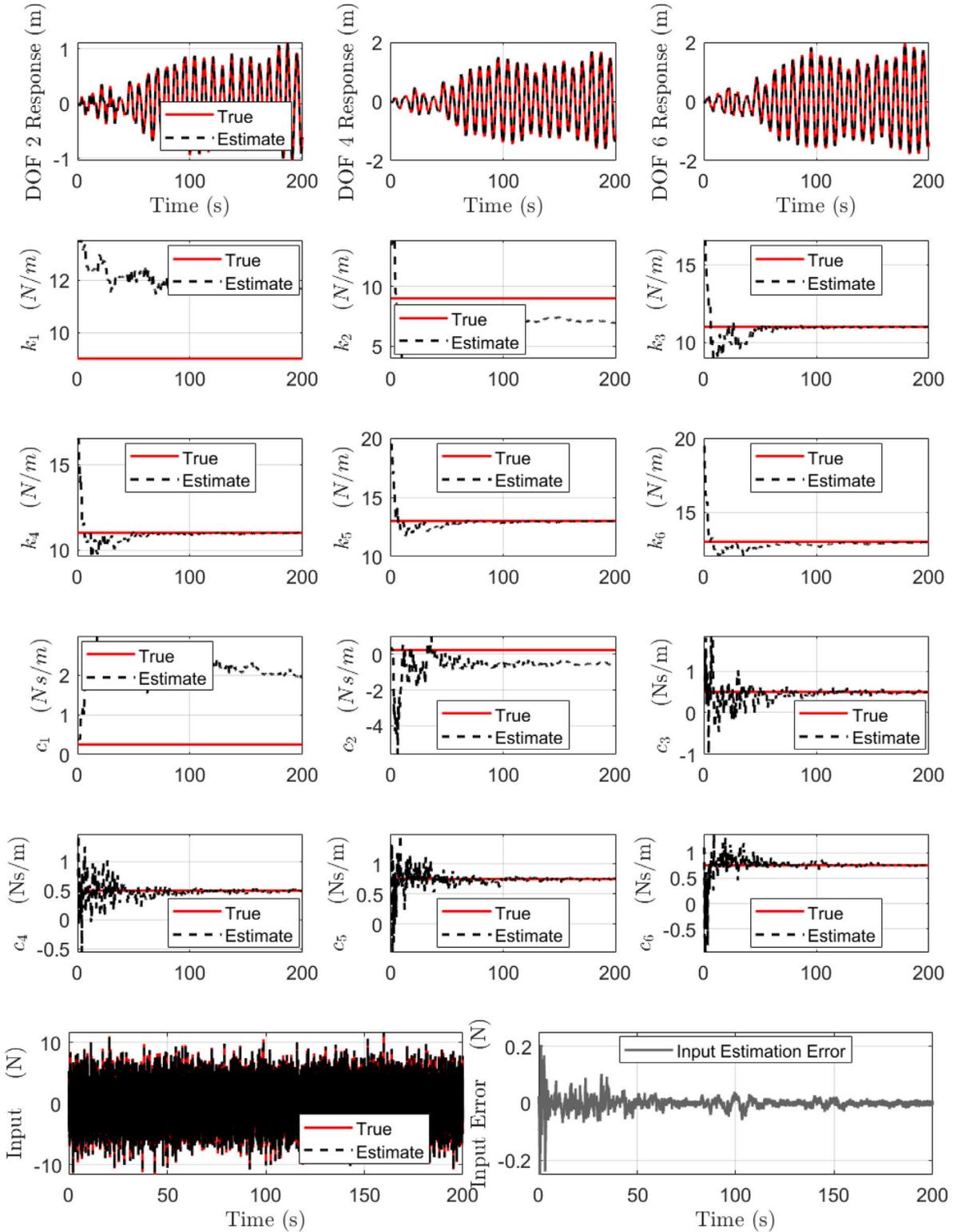

**Fig. 16.** Detailed parameter estimation and input error results for the system of Fig. 8 when the initial parameter set overestimate their value by 50%.

To clarify this point, Figs. 14–16 are provided. They show for the 6-DOF linear system with 4 sensors, and specifically, the true and estimated response (first row), the stiffness parameters (second and third row), the damping parameters (fourth and fifth row), and the input estimation and its error at DOF 6 (last row).

Fig. 14 refers to the case where the initial parameter set underestimate their value by 50%, Fig. 15 refers to the case where the initial parameter set underestimate their value by 25%, and Fig. 16 refers to the case where the initial parameter set overestimate their value by 50%. At this point, it is shown the parameter estimation for each one of them, and especially it is demonstrated that Fig. 15 provides the least input and parameter estimation error. The identification performance for each scenario is also shown.

Another concern is related to the noise parameters are their manual tuning and calibration. The trial-and-error approach could yield reasonable results for numerical studies where the actual values of the unknown quantities are known. The contribution of this study is to minimize this drawback using the Kullback–Leibler divergence which shows to provide the best initial parameter set without the user's interference. Notably, the application of this study to practical problems is required as a future research, justified and demonstrated through experimental examples with real data. However, here, an extensive numerical investigation was performed to propose the discussed approach and to provide insight into the correct directions. The Kalman filters types, importantly, have been verified to give accurate results in many system identification problems [63–65].

Specifically for the user-defined algorithmic parameters which affect the success of the input–parameter–state estimation, an investigation is provided for the RKF case, while for the UKF the reader is referred to [56]. The parameter $\mathbf{Q}$ is shown here to be the least important one in this methodology compared to the rest. Its role is to model the process error and its contribution seems to be directly related to the parameter $\mathbf{R}$; their numerical distance is of interest. Therefore, without a further investigation, $\mathbf{Q}$ is replaced by the unit matrix of proper dimensions. The parameter $\mathbf{R}$, though, is a crucial one when it comes to the correct dynamic state estimation. Its role is to model the measurement error (in this case, the pseudo-measurement error) and its contribution is to prevent the algorithm from diverging from the pseudo-measurements. The pseudo-measurements may be polluted, but they are an excellent guide for the behavior of the system. Importantly, having correctly estimated dynamic states is a primary requirement for the final correct convergence.

The parameter $\lambda^2$, for the RKF, is the most important algorithmic parameter when it comes to the performance of the system parameter estimation. The lower the $\lambda^2$ (very close to zero) the higher the possibility that the dynamic states are mis-estimated and critically, the optimal Kalman gain $\mathbf{J}$ to be close to singular or badly scaled. When this value is very high, the identification is far from the real system and a high divergence results. It is often convenient to set the regularization parameter constant and retain this value until the final step. For linear structural-mechanical systems the value is explored in the vicinity of $[0.01, 0.2]$. In this work, a $\lambda^2$ value close to 0.05 works adequately for the full input–parameter–state estimation [66–69]. The behavior of $\lambda^2$ with respect to the estimation error follows a L-curve shape [70,71]. The curve is approximately vertical for $\lambda^2 < \lambda^2_{optimal}$, and later becomes horizontal when $\lambda^2 > \lambda^2_{optimal}$, with a corner or a lower flat level near the optimal regularization parameter $\lambda^2_{optimal}$. The optimal value of the regularization parameter corresponds to the balance between confidence in the pseudo-measurements and the estimated quantities.

The final algorithmic parameter is the exponential parameter $\mu$; a very important one when it comes to the convergence duration. When this parameter is very high the final convergence of the system parameters happens within the first steps before the filter identifies the true values. When it is very low, the system parameters do not converge to single values. Values close to $10^{-2}$ seems a valid selection here, unless a possible damage is expected in real time with the identification.

Fig. 17 (four top plots) shows the relation between the norm of the residual of the system model estimation at the final step $\|r\|_2$ versus the $\mathbf{R}$ parameter. This demonstration is shown for the 6-DOF system without two of the six sensors. Here, as $\mathbf{R}$ is getting lower, the value of $\|r\|_2$ is also getting lower. This has a limit though. After that, a lower value of $\mathbf{R}$ does not provide a better performance. A recommended value of $\mathbf{R}$ is $10^{-10}$ times the unit matrix of proper dimensions. It also shows the relation between the norm of the residual of the system model estimation at the final step versus the value of $\lambda^2$ parameter. Subsequently, the relation is shown between the norm of the parameter error at the final step versus the value of $\lambda^2$ parameter. Here, the vertical and horizontal lines are clear before and after the optimal value of $\lambda^2$. For a chosen value close to but not $\lambda^2_{optimal}$, the methodology does not collapse; the convergence is still acceptable. A value far from the optimal though harms the procedure, regardless of how the rest of the algorithmic parameters are calibrated.

Finally, Fig. 17 (six bottom plots) shows the stiffness and damping parameter estimation when all algorithmic parameters remain the same except for $\mu$. The demonstration is shown for the 6-DOF system without two of the six sensors. The $\mu$ values considered are $10^{-1}$, $10^{-2}$, and $10^{-3}$. In the first case, the methodology does not have an adequate time to converge to the correct parameters. In the second case, the methodology converges to the correct parameters in a reasonable time without allowing some later polluted pseudo-measurements to re-estimate the already correctly estimated quantities. In the final case, the value of $\mu$ is so low that the methodology does not converge; a trend exists towards the correct parameter value, but polluted pseudo-measurements keep the filter from converging to a final estimate. Overall, it is beneficial to run the algorithm in parallel with different $\mu$ values, reject the ones that do not fluctuate after the initial learning phase, and choose the one which stops fluctuate after a critical time period.

Regarding the effect of taking the unmeasured acceleration responses equal to their estimated priors, an investigated has been provided in [50] for an insight into the convergence and stability of the estimates. It has been concluded that the results for the dynamic states and the input are acceptably satisfactory for a low number of sensors but not all parameters converge to the true values; especially the damping parameters where, by their nature, are excited the least. For an even lower number of sensors, the estimation is even poorer as expected. Although the dynamic states and the input are correctly estimated, almost all parameters diverge; this is an indication of a non-identifiable system for this combination of measurements and input.

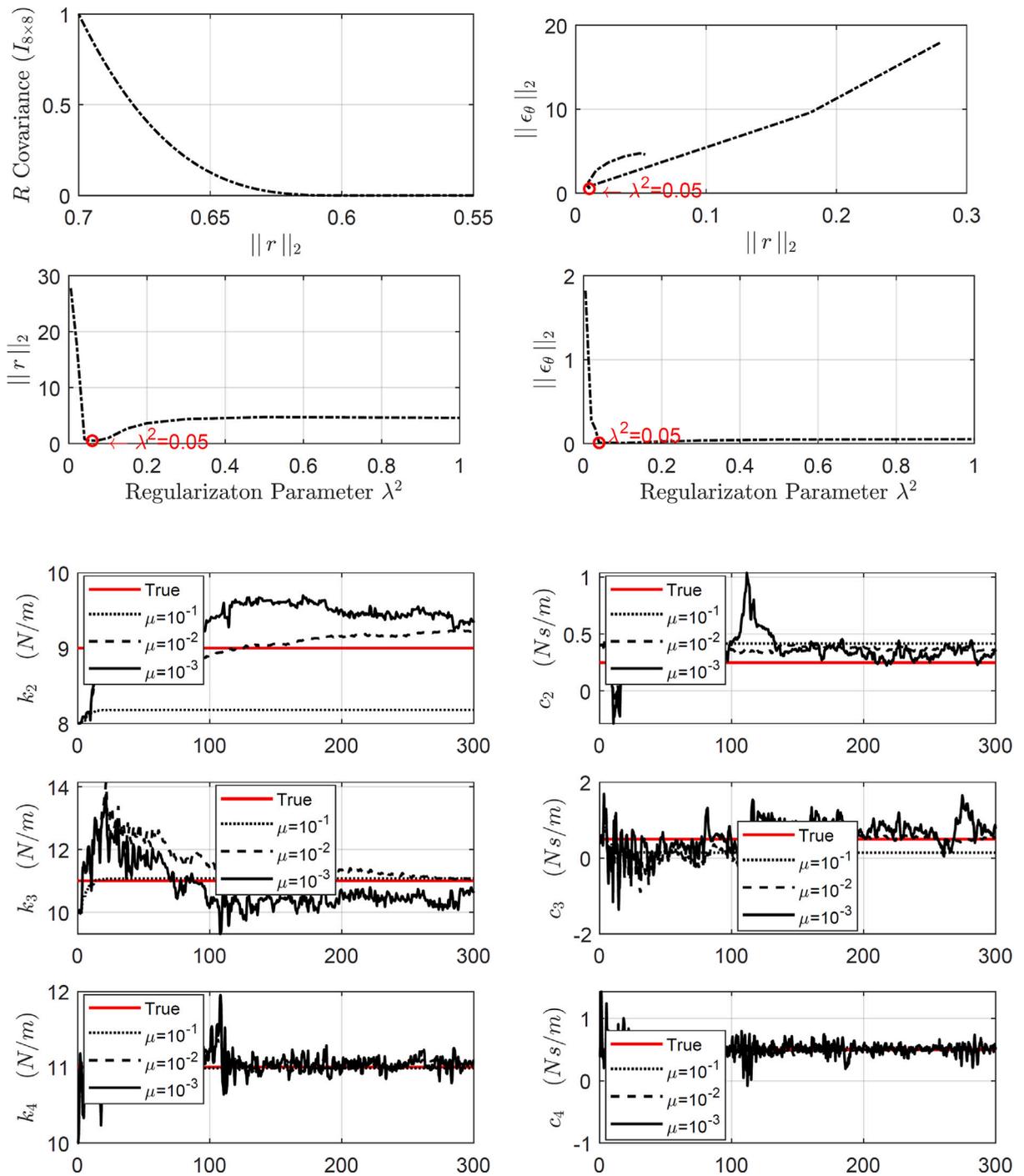

**Fig. 17.** It is shown (four top plots): $\mathbf{R}_d$ parameter versus the residual of the system model estimation at the final step $\|r\|_2$, optimal $\lambda^2$ parameter versus the residual of the system model estimation at the final step $\|r\|_2$ and the residual of the measurement and prediction error at the final step $\|\epsilon_\theta\|_2$, and (six bottom plots) $\mu$ scaling parameter investigation for the 6-DOF system of Fig. 8.

Regarding the justification of the method placed here, it seems that when using the Kullback–Leibler divergence for restricting the parametric distributions close to the initial assumption, one virtually forces these parameters to remain similar to the original initial assumption. This more or less would reflect a filter that does not need to do parameter estimation in the loop, but one which runs for a fixed assumed set of parameters. Therefore, one may misinterpret that the equivalent approach to what is shown here is running various filters of different samples of parameters in parallel and selecting the one corresponding to the best initial set. In reality though, the methodology is shown to automatically select the best initial parameter set which is unknown to the user. Without this approach, the algorithm would have converged to suboptimal results which, although partially reproduce the measured data, they do not provide the correct parameter estimates. More importantly, this approach does not force the posterior to be close to an assumed biased prior, but selects on its own the best prior assumptions. The need is derived from the fact that the filters do not always have the capacity to recover better fitting estimates when the initial parameter set are way off for input–parameter–state estimation problems.

Alternative approaches for future investigation may be searched in the Particle Filters, or Rao-Blackwellized filters [72,73] to verify the general applicability of the Kullback–Leibler divergence. However, so far it has not been reported an input–parameter–state estimation methodology based on those.

Importantly, the current approach describes the Kullback–Leibler divergence to be estimated for every time step, and not for the whole analysis cumulatively. It is performed in every time step of the filter computation as implied by all plots and it is used in the sequential updating of the parameters compared always to the initial distributions. As a result, it is not that only the final value is simply observed in order to choose a best initial parameter guess.

A severe limitation of the structural health monitoring applications is the need of using a low number of sensors, and potentially create a pseudo-measurements approach from acceleration data. This is realistic, though, and in fact the main use of these filters is to alleviate the need from full field measurements and allow a realistic assumption of few and sparse sensors.

Finally, a deeper uncertainty quantification is recommended for future research showing results for many initial parameter sets and large parameter vector dimensions for large-scale system applications in order the suggested approach to be validated in a statistical manner.

## 10. Conclusions

The capability of a novel Kullback–Leibler divergence method was examined herein within the Kalman filter framework to select the input–parameter–state estimation execution with the most plausible results. This identification suffers from the uncertainty related to obtaining different results from different initial parameter set guesses, and the examined approach used the information gained from the data in going from the prior to the posterior distribution to address the issue. Firstly, the Kalman filter was performed for a number of different initial parameter sets providing the system input–parameter–state estimation. Secondly, the resulting posterior distributions were compared simultaneously to the initial prior distributions using the Kullback–Leibler divergence. Finally, the identification with the least Kullback–Leibler divergence was selected as the one with the most plausible results.

Overall, this method allowed for the joint input–parameter–state with:

1. Different initial parameter set considerations.
2. Real time evaluation of the posterior distributions.
3. Simultaneous and direct comparison of different identification executions, in the same test.
4. Automatic selection of the identification with the most plausible results.
5. Independent to the system type application.

Importantly, the method was shown to select the better performed identification in linear, nonlinear, and limited information applications, providing a powerful tool for system monitoring.

## CRediT authorship contribution statement

**Marios Impraimakis:** Conceptualization, Methodology, Software, Validation, Formal analysis, Investigation, Data curation, Visualization, Writing – original draft, Writing – review & editing.

## Declaration of competing interest

The author declares that there is no conflict of interest.

## Data availability

Data will be made available on request.

## Acknowledgments

The author would like to gratefully acknowledge the reviewers for their constructive comments, and Andrew W. Smyth for the previous insightful discussions on the topic.